\documentclass[fleqn,usenatbib]{mnras}

\usepackage{newtxtext,newtxmath}
\usepackage{hyperref}

\usepackage[T1]{fontenc}

\DeclareRobustCommand{\VAN}[3]{#2}
\let\VANthebibliography\thebibliography
\def\thebibliography{\DeclareRobustCommand{\VAN}[3]{##3}\VANthebibliography}

\newcommand{\msun}{\ensuremath{{\mathrm{M}_{\odot}}}}

\usepackage{graphicx}	
\usepackage{amsmath}	
\usepackage{orcidlink}
\usepackage{lipsum}
\usepackage{widetext}
\newcommand*{\mailto}[1]{\href{mailto:#1}{#1}}

\title[Constrained merger trees]{Growing the first galaxies' merger trees}

\author[E.~O.~Nadler et al.]{
Ethan O.~Nadler$^{\orcidlink{0000-0002-1182-3825}1,2}$\thanks{E-mail: \mailto{enadler@carnegiescience.edu}},
Andrew Benson,$^{\orcidlink{0000-0001-5501-6008}1}$\thanks{E-mail: \mailto{abenson@carnegiescience.edu}}
Trey Driskell,$^{\orcidlink{0000-0001-9472-7179}2}$
Xiaolong Du,$^{\orcidlink{0000-0003-0728-2533}1,3}$,
and Vera Gluscevic$^{\orcidlink{0000-0002-3589-8637}2}$
\\
$^{1}$Carnegie Observatories, 813 Santa Barbara Street, Pasadena, CA 91101, USA\\
$^{2}$Department of Physics $\&$ Astronomy, University of Southern California, Los Angeles, CA, 90007, USA\\
$^{3}$Department of Physics and Astronomy, University of California, Los Angeles, 430 Portola Plaza, Los Angeles, CA 90095, USA\\
}

\pubyear{2022}

\begin{document}
\label{firstpage}
\pagerange{\pageref{firstpage}--\pageref{lastpage}}
\maketitle

\begin{abstract} 
Modelling the growth histories of specific galaxies often involves generating the entire population of objects that arise in a given cosmology and selecting systems with appropriate properties. This approach is highly inefficient when targeting rare systems such as the extremely luminous high-redshift galaxy candidates detected by JWST. Here, we present a novel framework for generating merger trees with branches that are guaranteed to achieve a desired halo mass at a chosen redshift. This method augments extended Press Schechter theory solutions with constrained random processes known as Brownian bridges and is implemented in the open-source semi-analytic model \textsc{Galacticus}.
We generate ensembles of constrained merger trees to predict the growth histories of seven high-redshift JWST galaxy candidates, finding that these systems most likely merge $\approx 2~\mathrm{Gyr}$ after the observation epoch and occupy haloes of mass $\gtrsim 10^{14}~\msun$ today. These calculations are thousands of times more efficient than existing methods, are analytically controlled, and provide physical insights into the evolution of haloes with rapid early growth. Our constrained merger tree implementation is publicly available at \url{https://github.com/galacticusorg/galacticus}.
\end{abstract}




\section{Introduction}

\begin{figure*}
    \centering
    \includegraphics[width=\textwidth]{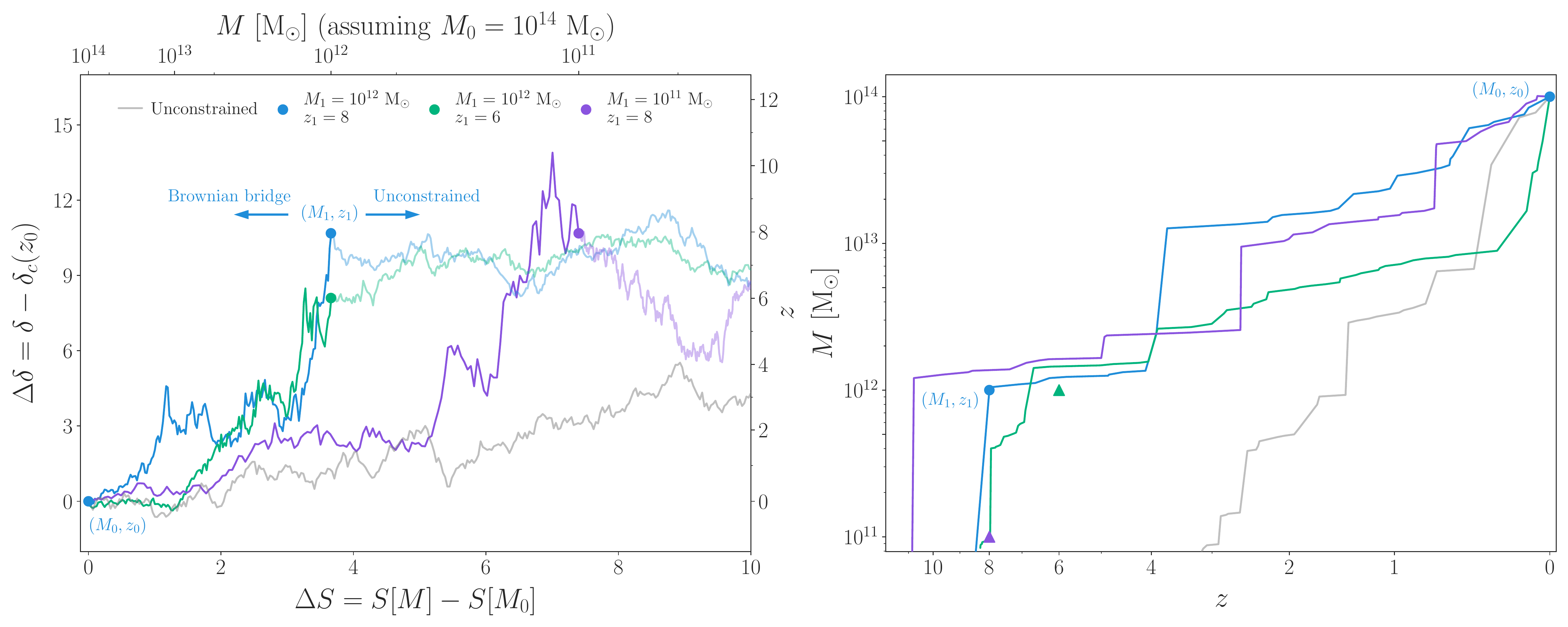}
    \vspace{-4mm}
    \caption{\emph{Left}: Constrained excursions (coloured lines) and a standard unconstrained excursion (grey line), that represent possible growth histories of a final halo of mass $M_0=10^{14}~M_{\mathrm{\odot}}$ at redshift $z_0=0$. The constrained Brownian bridge excursions are forced to pass through an overdensity and mass variance corresponding to a halo of mass $M_1 = 10^{12}~\msun$ at $z_1=8$ (blue), $M_1 = 10^{12}~\msun$ at $z_1=6$ (green), and $M_1 = 10^{11}~\msun$ at $z_1=8$ (purple). After each constraint is achieved, growth histories are modeled with unconstrained excursions until $z_0$. $M_0$, $z_0$, $M_1$, and $z_1$ are free input parameters in our constrained solutions for any $M_1<M_0$ and $z_1>z_0$. Although excursions appear non-monotonic on the redshift axis, only their successive maxima correspond to steps in halo growth (shown in the right-hand panel). \emph{Right}: Corresponding halo growth histories. The purple and green points from the left-hand panel are shown as lower limits on halo mass at their respective constrained masses and redshifts because these excursions first cross above the corresponding critical overdensities at larger mass scales. However, the constrained merger tree algorithm we present yields growth histories that \emph{precisely} satisfy desired constraints, like the blue line in the right-hand panel.}
    \label{fig:random_walk}
\end{figure*}

The extended Press--Schechter (ePS) formalism \citep{Press1974,Bond1991,Bower1991,Lacey1993} provides an analytic recipe for predicting the formation histories of dark matter haloes, both in standard $\Lambda$CDM cosmologies (e.g., \citealt{Sheth9907024,Giocoli0611221}) and beyond (e.g., \citealt{Benson12093018}). The ePS formalism is based on the set of excursions $\delta(S)$ that a point in the linear overdensity field executes as it is smoothed on successively smaller scales, commonly paramterized by the mass variance $S$ (see Fig.~\ref{fig:random_walk} for examples). Mergers in a halo's growth history correspond to successive maxima of $\delta(S)$; the masses and redshifts of these mergers are determined by the values of the mass variance at which the linear overdensity reaches the threshold necessary for halo formation.

Models that leverage the ePS formalism have been calibrated to numerical simulations and are remarkably useful for rapidly generating ensembles of halo growth histories and merger trees (for a review, see \citealt{Zentner0611454} and references therein). However, existing ePS-based approaches are highly inefficient when modelling specific astrophysical systems. In particular, most ePS algorithms model excursions as uncorrelated random walks and thus cannot guarantee that constraints on a halo's growth history beyond its present-day mass are satisfied. We will show that the ability to generate merger trees that satisfy specific constraints is desirable for modelling rare astrophysical systems such as the first detectable galaxies. In particular, we derive constrained excursion set solutions that allow us to directly generate the growth histories of rare high-redshift galaxies, rather than generating exponentially many realizations and selecting those with the desired properties. Our constrained merger tree realizations of the first galaxies are up to $\sim 10^4$ times more efficient than standard ePS calculations for sampling mean growth histories, and up to~$\sim 10^8$ times faster for sampling outliers among this population.

Mathematically, excursions are usually modeled as uncorrelated random walks with normally distributed transition probabilities under the usual assumption of a Gaussian linear overdensity field.\footnote{These assumptions imply that standard ePS excursions are Wiener processes in the continuum limit; thus, the constrained solutions we present can be applied to any stochastic process $y(x)$ with a suitable barrier.} However, as emphasized by \cite{Zentner0611454}, the uncorrelated nature of ePS excursions is not a prediction of the theory; rather, it is a simplifying assumption that may not hold in practice. For example, any window function other than the sharp $k$-space filter correlates Fourier modes of the density field \citep{Bond1991}, and haloes' large-scale environments also introduce correlations in their merger histories. Here, we go beyond the usual ePS assumption of uncorrelated random walks by implementing constrained excursions of the linear overdensity field in a merger tree algorithm. To achieve this, we use constrained random processes known as Brownian bridges. Unlike uncorrelated random walks, Brownian bridges are \emph{guaranteed} to pass through desired points along their trajectories.

Physically, our Brownian bridge solutions allow us to fix a point in a halo's growth history at a chosen mass scale below its final mass and construct excursions on larger mass scales that are consistent with this constraint. This implies that a branch of a final halo's merger tree must exceed a desired mass by a specific redshift. In detail, we derive halo merger rates for excursions that begin at $(\delta_0,S_0)$ and are \emph{guaranteed} to pass through a specific point $(\delta_1,S_1)$ for any $S_1>S_0$ and $\delta_1>\delta_0$. We implement these merger rates in a new merger tree algorithm, which guarantees that a halo with final mass $M_0$ at redshift $z_0$ has a progenitor of exactly mass $M_1$ at redshift $z_1$ for \emph{any} $M_1<M_0$ and~$z_1>z_0$. All four of these quantities (i.e., $M_0$, $z_0$, $M_1$, and $z_1$) are free input parameters in our approach. Thus, halo growth histories generated using our constrained merger tree algorithm may be \emph{arbitrarily rare} relative to the unconstrained distribution for a final halo of mass $M_0$ at redshift $z_0$. Crucially, the distribution of these constrained trajectories is precisely equal to the subset of unconstrained excursions that satisfy the desired constraints; our predictions are therefore statistically unbiased, even for extremely rare systems.

The left-hand panel of Fig.~\ref{fig:random_walk} shows examples of constrained and unconstrained excursions that represent a final halo of mass $M_0=10^{14}~\msun$ at $z_0=0$; the right-hand panel shows the corresponding halo growth histories. More extreme constrained solutions -- e.g., the blue model that exceeds $M_1=10^{12}~\msun$ at $z_1=8$, making it an $\approx 4\sigma$ outlier relative to the underlying halo population at that redshift -- reach large overdensities more quickly along their trajectories. Because halo growth can be read off as the successive maxima of an excursion, followed from left to right on the left-hand panel of Fig.\ \ref{fig:random_walk}, such extreme constraints yield rapid early growth before $z_1$ followed by relatively quiescent merger histories. On the other hand, standard unconstrained ePS solutions, like the grey excursion in Fig.~\ref{fig:random_walk}, often yield significant major mergers at late times for this final halo mass, as reflected in the corresponding growth history.

We leverage this ability to generate rare unbiased halo growth histories to grow the first large ensembles of merger trees representing the earliest detectable galaxies; this task would otherwise be computationally intractable. Due to observational selection effects, observed high-redshift galaxies are generally the most luminous objects that exist at early times, which implies that they likely occupy the most rapidly-growing haloes at these epochs (e.g., \citealt{Mason220714008}). For example, surprisingly luminous galaxies at redshifts $z\gtrsim 10$ have been identified in Hubble Space Telescope (\emph{HST}; \citealt{Oesch160300461}) and Hyper-Suprime-Cam Subaru Strategic Program (HSC-SSP; \citealt{Harikane211209141}) data. Recently, similarly remarkable galaxy candidates have been detected in JWST data (e.g., \citealt{Labbe220712446,Naidu220709434,Robertson221204480,Curtis-Lake221204568}) data, though the nature of these systems is still highly uncertain due to instrumental calibration (e.g., \citealt{Boyer220903348}), dust, initial-mass function, and nebular-emission modelling (e.g., \citealt{Zavala220801816,Naidu220802794}), as well as limited spectroscopic follow-up.

Estimates based on abundance matching and extreme value statistics suggest that the extremely luminous high-redshift galaxy candidates detected by JWST occupy haloes with masses $\gtrsim 10^{11}$ at $z\gtrsim 8$, implying that they are at least $\approx 3\sigma$ outliers relative to a standard $\Lambda$CDM halo population (e.g., \citealt{Lovell220810479}); as a result, the very existence of these early, luminous objects may pose challenges to $\Lambda$CDM itself \citep{Boylan-Kolchin220801611}. Our results allow us to statistically predict the merger rates and descendant mass distributions of these early galaxies, assuming they are hosted by rare haloes in a standard $\Lambda$CDM cosmology. Note that, we do not attempt to forward model the baryonic components of rare early haloes in this paper; however, we intend to explore such modelling in future work.

Beyond the high-redshift applications considered in this paper, we anticipate that constrained excursion set solutions will generally be useful for modelling other kinds of astrophysical systems. For example, near-field cosmological observations provide detailed constraints on the merger histories of systems like the Milky Way (e.g., see \citealt{Helmi200204340} for a review). Such systems are also ``rare,'' in the sense that halo growth histories generated using standard ePS theory are very unlikely to match the specific set of relevant constraints. Thus, even if a system's growth history is not extreme relative to the mean of the underlying distribution, our algorithm yields merger trees that are guaranteed to match its specific properties and sample the remaining scatter in possible growth histories. We set the stage for such applications in this work by demonstrating that certain choices of our Brownian bridge constraint parameters yield merger events that resemble the infall of the Large Magellanic Cloud into the Milky Way. We also intend to explore other classes of constrained excursion set solutions, along with relevant astrophysical and cosmological applications, in future studies.

This paper is organized as follows. In Section~\ref{sec:ePS_theory}, we review essential ingredients of excursion set theory and derive the first-crossing distribution for Brownian bridge excursions; in Section~\ref{sec:implementation}, we describe our method for constructing constrained merger trees based on these first-crossing distributions; in Section~\ref{sec:results}, we present halo growth histories, formation time distributions, and merger rates for constrained merger trees; in Section~\ref{sec:applications}, we use our technique to predict the merger rates and descendants of high-redshift JWST galaxy candidates; in Section~\ref{sec:mw_lmc}, we model the Milky Way--Large Magellanic Cloud merger; we address potential caveats in Section~\ref{sec:calibration}, and we conclude in Section~\ref{sec:conclusions}. We also present several Appendices: in Appendix \ref{sec:integral_equations}, we present the full integral equations that determine our unconstrained and constrained first-crossing rate distributions; in Appendix \ref{sec:convergence_tests}, we show that our constrained solutions converge to the unconstrained distribution of halo growth histories in appropriate limits, and we demonstrate the numerical convergence of our constrained results; in Appendix~\ref{sec:descendant_derivation}, we derive the distribution of descendant halo masses that enters our high-redshift galaxy calculations, given the $z=0$ halo mass function.

Throughout, we adopt the best-fit \emph{Planck} cosmology with $H_0=67.66~\mathrm{km}~\mathrm{s}^{-1}~\mathrm{Mpc}^{-1}$, $\Omega_m = 0.3111$, $\Omega_b=0.049$, $\Omega_{\Lambda}=0.6889$, $n_s=0.9665$, and $\sigma_8=0.8102$ \citep{Planck2018}. Note that, because we construct dark matter-only halo growth histories in this paper, we set $\Omega_\mathrm{b}=0$ when computing halo masses. Throughout, $\log$ denotes the base-$10$ logarithm.

\section{Excursion Set Theory}
\label{sec:ePS_theory}

In this section, we describe our excursion set theory procedure for modelling constrained dark matter halo growth histories. In Section~\ref{sec:ingredients}, we briefly review standard ePS theory ingredients, including an expression for the first-crossing distribution of excursions given the collapse barrier. We use this expression to derive the first-crossing distribution for standard unconstrained excursions in Section~\ref{sec:unconstrained_solution} and our constrained Brownian bridge excursions in Section~\ref{sec:contrained_solution}. We present examples of the resulting first-crossing rate distributions in Section~\ref{sec:first-crossing_examples}.

\subsection{Standard ingredients}
\label{sec:ingredients}

The first ingredient is the linear power spectrum of matter density fluctuations $P(k)$, which we use to define the mass variance $\sigma(M)\equiv \sqrt{S(M)}$ (i.e., the fractional root-variance of the linear overdensity field extrapolated to $z = 0$), according to
\begin{equation}
    S(M)\equiv \sigma^2(M) = \frac{1}{2\pi^2}\int_{0}^{\infty}4\pi k^2 P(k) W^2(k|M) \mathrm{d}k,\label{eq:root-variance}
\end{equation}
where $k$ is the comoving wavenumber and $W(k|M)$ is a window function that smooths the linear overdensity field $\delta$ on a mass scale $M$. We assume a sharp $k$-space window function
\begin{equation}
    W_k(k|M) = \begin{cases} 1 &k\leq k_S(M)\\ 0 &k>k_S(M),
\end{cases}
\end{equation}
where the subscript $S$ indicates that the halo mass $M$ is evaluated relative to the root-variance defined by equation (\ref{eq:root-variance}). In standard ePS theory and in all of our calculations, steps along the trajectory $\delta(S)$ correspond to draws from a Gaussian random field in Fourier space and are thus uncorrelated for a sharp $k$-space window function. However, to compute $\sigma(M)$, we follow \cite{Benson12093018} by using a real-space top-hat window function,
\begin{equation}
    W_R(k|M) = \frac{3[\sin(kR)-kR\cos(kR)]}{(kR)^3},
\end{equation}
where the filtering scale $R$ is related to halo mass $M$ via
\begin{equation}
    R = \left(\frac{3M}{4\pi\bar{\rho}}\right)^{1/3},
\end{equation}
and $\bar{\rho}$ is the mean matter density at $z=0$ \citep{Lacey1993}. We use a real-space top-hat window function because it allows for this unambiguous mass definition and corresponds to a physically intuitive scenario in which localized overdensities collapse into haloes.

The next ingredient is the critical threshold overdensity $\delta_c(z)$, which represents a condition for the gravitational collapse of a density perturbation in linear theory. This quantity acts as an absorbing barrier $B(S)$, because trajectories that first up-cross the barrier at variance $S$ collapse into a halo of mass $M[S]$.\footnote{We write $M[S]$ as a functional because of the integral relation between these variables in equation (\ref{eq:root-variance}).} We follow \cite{Bond1991} by assuming that
\begin{equation}
B(S,t) = \delta_c(M[S],t)=\frac{\delta_{c,0}}{D(t)},\label{eq:barrier}
\end{equation}
where $\delta_\mathrm{c,0}\approx 1.686$ is the linear-theory collapse threshold and $D(t)$ is the linear growth factor \citep{Carroll1992}. In particular, following \cite{Benson12093018}, we use the barrier in equation (\ref{eq:barrier}) rather than the `remapped' barrier introduced in \cite{Sheth9907024} that is calibrated to match halo mass functions from cosmological simulations because we only aim to compute merger rates rather than halo mass functions. 

To predict the progenitors of a halo of mass $M_0$ at redshift $z_0$, we need to compute the conditional distribution of halo masses that would lead to this final mass (i.e., the conditional mass function; \citealt{Lacey1993}). In the excursion set approach, this corresponds to solving for the first up-crossing of $\delta(S)$ given excursions that begin from the point $(\delta_0,S_0)$ corresponding to the final halo in question. As described in \cite{Benson12093018}, computing merger rates as a function of the relative variance,
\begin{equation}
\tilde{S}\equiv S-S_0,
\end{equation}
amounts to solving the first-crossing problem with a modified barrier \begin{equation}
\tilde{B}(\tilde{S},t',t_0)=B(\tilde{S}+S_0,t')-B(S_0,t_0),
\end{equation} 
where $t_0$ is defined implicitly (given $\delta_0$) via equation (\ref{eq:barrier}) and $t'<t$ is an earlier time in this halo's history. We then estimate the first-crossing rate using the finite-difference method from \cite{Benson12093018}. In particular, we set $t'=(1-\epsilon)t_0$, where $\epsilon\ll 1$, which implies that the first-crossing rate of the effective barrier $\tilde{B}(\tilde{S},t',t_0)$ is then $\mathrm{d}f/\mathrm{d}t=f(S)/\epsilon t_0$, where $f(S)\mathrm{d}S$ is the fraction of excursions that first up-cross the barrier between $S$ and $S + \mathrm{d}S$. 

Thus, the \emph{first-crossing rate distribution} $\mathrm{d}f(S)/\mathrm{d}t$ gives the rate at which excursions first achieve the critical overdensity, and physically corresponds to the branching rate of dark matter haloes into lower-mass progenitors. The problem of constructing halo merger histories therefore boils down to finding the first-crossing distribution $f(S)$. \cite{Zhang0508384} formulated a generic integral equation for $f(S)$,
\begin{equation}
    1 = \int_0^S f(\hat{S})\mathrm{d}\hat{S} + \int_{-\infty}^{\tilde{B}(S)} P(\delta,S)\mathrm{d}\delta,\label{eq:f_integral}
\end{equation}
where $P(\delta,S)\mathrm{d}\delta$ is the probability that a trajectory lies between $\delta$ and $\delta + \mathrm{d}\delta$ at variance $S$, having never crossed the barrier at any smaller $S$. Equation (\ref{eq:f_integral}) follows from mass conservation, because all trajectories must either ($i$) cross $B(S)$ at $0<\hat{S}<S$, or ($ii$) lie below the barrier at $S$, having never crossed it at smaller $S$ \citep{Benson12093018}. $P(\delta,S)$ can be further decomposed according to
\begin{equation}
    P(\delta,S) = P_0(\delta,S) - \int_{0}^S f(\hat{S})P_0[\delta,S|\tilde{B}(\hat{S}),\hat{S}]\mathrm{d}\hat{S},\label{eq:probability_expansion}
\end{equation}
where the first term is the distribution of trajectories in the absence of a barrier, and the second term is the fraction of trajectories that first up-crossed the barrier at $\hat{S}<S$ and then reach $(\delta,S)$. Combining equations (\ref{eq:f_integral}) and (\ref{eq:probability_expansion}) yields
\begin{align}
    1 = & \int_0^S f(\hat{S})\mathrm{d}\hat{S} & \nonumber \\ + & \int_{-\infty}^{\tilde{B}(S)} \left[P_0(\delta,S) - \int_{0}^S f(\hat{S})P_0[\delta,S|\tilde{B}(\hat{S}),\hat{S}]\mathrm{d}\hat{S}\right]\mathrm{d}\delta. \label{eq:f_integral_final} &
\end{align}
We follow the strategy for numerically solving equation (\ref{eq:f_integral_final}) by discretizing the integrals over grids of $\hat{S}$, as described in Appendix A of \cite{Benson12093018}. Thus, in this approach, we must specify $P_0(\delta,S)$ and $P_0[\delta,S|\tilde{B}(\hat{S}),\hat{S}]$ to define an excursion set solution.

\subsection{Unconstrained excursions}
\label{sec:unconstrained_solution}

In the standard excursion set problem, the distribution of trajectories $P_0(\delta,S)$ in the absence of a barrier (i.e., the first term in equation (\ref{eq:probability_expansion})) is a normal distribution with zero mean and variance $S$,
\begin{equation}
    P_0(\delta,S) = \mathcal{N}(\delta,S) = \frac{1}{\sqrt{2\pi S}}\exp\left(-\frac{\delta^2}{2S}\right).\label{eq:unconstrained_p}
\end{equation}
Next, we consider trajectories that start at $(\tilde{B}(\hat{S}),\hat{S})$ and end at $(\delta,S)$ to derive the distribution inside the integrand of the second term in equation (\ref{eq:probability_expansion}). For uncorrelated random walks, there is no net drift in the trajectories, so the effective difference in $\delta$ between these two points is simply 
\begin{equation}
\Delta \delta = \delta - \tilde{B}(\hat{S}).
\end{equation}
The covariance between these two points is $\mathrm{Cov}(\hat{S},S) = \min(\hat{S},S) = \hat{S}$ (since $\hat{S}<S$ by construction). Thus, the residual variance is
\begin{equation}
    \Delta S = S - \mathrm{Cov}(\hat{S},S) = S - \hat{S}.
\end{equation}
The desired conditional distribution function is therefore 
\begin{equation}
    P_0[\delta,S|\tilde{B}(\hat{S}),\hat{S}] = \mathcal{N}(\Delta \delta,\Delta S) = \frac{1}{\sqrt{2\pi (S-\hat{S})}}\exp\left(-\frac{(\delta-\tilde{B}(\hat{S}))^2}{2(S-\hat{S})}\right).\label{eq:unconstrained_relative_p}
\end{equation}

Plugging equations (\ref{eq:unconstrained_p}) and (\ref{eq:unconstrained_relative_p}) into equation (\ref{eq:f_integral_final}) and evaluating the integral in the second term over $\mathrm{d}\delta$ analytically yields the final form of our integral equation for $f(S)$. This equation is presented explicitly in Appendix \ref{sec:unconstrained_derivation}. We solve for $f(S)$ following the numerical scheme presented in \cite{Benson12093018}, using the midpoint integration method from \cite{Du160820575}. We then use the resulting first-crossing distributions to generate unconstrained halo growth histories as described in Section~\ref{sec:merger_tree_construction}. Two free quantities specify the distribution of unconstrained halo growth histories:
\begin{enumerate}
    \item $M_0$, the final halo mass, which enters equation (\ref{eq:f_integral_final}) through the definition $\tilde{S}=S-S_0$ and through the effective barrier $\tilde{B}(S)$ implicitly, via $\delta_c(M[S],t)$;
    \item $z_0$, the redshift at which the final halo exists, which only enters equation (\ref{eq:f_integral_final}) via the dependence of the effective barrier on $\delta_c(M[S],t)$. The final redshift $z_0$ need not equal zero.
\end{enumerate}

\subsection{Constrained Brownian bridge excursions}
\label{sec:contrained_solution}

We construct constrained merger trees with branches that begin at $(\delta_0,S_0)$ and are forced to pass through the point $(\delta_1,S_1)$ using the Brownian bridge first-crossing rate distribution for $S_0\leq S\leq S_1$, and then switch to an unconstrained solution for $S>S_1$. We justify this procedure in Section~\ref{sec:implementation}. Note that, Brownian bridge excursions may up-cross $\delta_1$ at $S<S_1$. Thus, such excursions formally place a lower limit on the mass of the progenitor on a single branch of the merger tree at that redshift; we refer to this as the ``constrained'' branch. Our implementation of Brownian bridge excursions in \textsc{Galacticus} exploits symmetries of the excursion set problem to guarantee that this constrained branch reaches a mass of precisely $M_1$ at redshift $z_1$. We discuss these points in detail below, and focus on the solution for the constrained first-crossing rate here.

For the purposes of computing the distributions that enter our solution for the first-crossing distribution in equation (\ref{eq:f_integral_final}), there are two main differences between the Brownian bridge and unconstrained cases. First, unlike uncorrelated random walks, Brownian bridge excursions have a net drift characterized by
\begin{equation}
    \mu(S) = \langle \delta(S) \rangle = \delta_0 +  \frac{S-S_0}{S_1-S_0}(\delta_1-\delta_0).
\end{equation}
Second, the covariance for two points $(\tilde{B}(\hat{S}),\hat{S})$ and $(\delta,S)$ along a Brownian bridge is
\begin{equation}
    \mathrm{Cov}(\tilde{B}(\hat{S}),\delta) = \frac{(S_1-S)(\hat{S}-S_0)}{S_1-S_0},\label{eq:drift_brownian_bridge}
\end{equation}
for $\hat{S}<S$. Note that, as $S\rightarrow S_1$, $\mu(S)\rightarrow \delta_1$ and $\mathrm{Cov}(\tilde{B}(\hat{S}),\delta)\rightarrow 0$, because we can recursively apply the Brownian bridge drift and covariance as we update $\delta$ and $S$ at each step along an excursion. In particular, any trajectory that starts at $(S_0,\delta_0)$ and passes through $(\hat{S},\hat{\delta})$ on its way to $(S_1,\delta_1)$, where $S_0<\hat{S}<S_1$, is itself a Brownian bridge that starts at $(\hat{S},\hat{\delta})$ and ends at $(S_1,\delta_1)$. Thus, steps of mean size $\delta_1$ are forced to occur as $S_1$ is approached, ensuring that the trajectories up-cross $\delta_1$ before $S_1$ is reached, as desired.

Next, to compute the effective difference in $\delta$ between the points $(\tilde{B}(\hat{S}),\hat{S})$ and $(\delta,S)$ for Brownian bridge excursions, we subtract the appropriate drift term from each point,
\begin{align}
    \Delta \delta' &= [\delta-\mu(S)] - [\tilde{B}(\hat{S})-\mu(\hat{S})]& \nonumber \\ &= \delta - \tilde{B}(\hat{S}) - \frac{S-\hat{S}}{S_1-S_0}(\delta_1-\delta_0).&\label{eq:relative_drift}
\end{align}
Similarly, the residual variance is given by
\begin{align}
    \Delta S' &= \mathrm{Var}(S) - \mathrm{Cov}(\tilde{B}(\hat{S}),\delta) &\nonumber \\ &= \frac{(S_1-S)(S-S_0)}{S_1-S_0} - \frac{(S_1-S)(\hat{S}-S_0)}{S_1-S_0}& \nonumber \\
    & = \frac{(S_1-S)(S-\hat{S})}{S_1-S_0}. \label{eq:cov_brownian_bridge} &
\end{align}
Note that, $\Delta \delta' \rightarrow \Delta \delta$ and $\Delta S' \rightarrow \Delta S$ in the limit $S_1 \rightarrow \infty$. Physically, this limit represents constrained trajectories that must exceed a very small halo mass at $z_1$, in which case the constrained first-crossing rate distribution equals that for the unconstrained case, as expected. 

With equations (\ref{eq:drift_brownian_bridge}) and (\ref{eq:cov_brownian_bridge}) in hand, we can now re-write the distribution functions that appear in equation (\ref{eq:probability_expansion}) for constrained Brownian bridge excursions. Specifically, the distribution of trajectories in the absence of a barrier becomes
\begin{align}
    &P'_0(\delta,S) = \mathcal{N}(\Delta \delta'|_{\hat{S}\rightarrow S_0},\Delta S'|_{\hat{S}\rightarrow S_0})& \nonumber \\
    & = \mathcal{N}\left(\delta - \tilde{B}(S_0) - \frac{S-S_0}{S_1-S_0}(\delta_1-\delta_0),\frac{(S_1-S)(S-S_0)}{S_1-S_0}\right). \label{eq:constrained_p} &
\end{align}
The conditional distribution function appearing in equation (\ref{eq:probability_expansion}) follows analogously, without taking the $\tilde{S}\rightarrow S_0$ limit in the expressions for $\Delta \hat{\delta}$ and $\Delta \hat{S}$
\begin{align}
    &P'_0[\delta,S|\tilde{B}(\hat{S}),\hat{S}] = \mathcal{N}(\Delta \delta',\Delta S')& \nonumber \\
    & = \mathcal{N}\left(\delta - \tilde{B}(\hat{S}) - \frac{S-\hat{S}}{S_1-S_0}(\delta_1-\delta_0),\frac{(S_1-S)(S-\hat{S})}{S_1-S_0}\right). \label{eq:constrained_relative_p} &
\end{align}

Plugging equations (\ref{eq:constrained_p}) and (\ref{eq:constrained_relative_p}) into equation (\ref{eq:f_integral_final}) and evaluating the integral in the second term over $\mathrm{d}\delta$ analytically yields the final form of our integral equation for $f(S)$ in the constrained case. This equation is presented explicitly in Appendix \ref{sec:constrained_derivation}, and we solve for $f(S)$ in the Brownian bridge case following the same numerical scheme as for the unconstrained problem. Four free quantities specify the distribution of constrained halo growth histories:
\begin{enumerate}
\item $M_0$, the final halo mass, which (as in the unconstrained case) enters equation (\ref{eq:f_integral_final}) through the definition $\tilde{S}=S-S_0$ and through the effective barrier $\tilde{B}(S)$ implicitly, via $\delta_c(M[S],t)$, and (only in the constrained case) enters equations (\ref{eq:constrained_p}) and (\ref{eq:constrained_relative_p}) via $S_0$; 
\item $z_0$, the redshift at which the final halo exists, which (as in the unconstrained case) enters equation (\ref{eq:f_integral_final}) via the dependence of the effective barrier on $\delta_c(M[S],t)$, and (only in the constrained case) enters equations (\ref{eq:constrained_p}) and (\ref{eq:constrained_relative_p}) via $\delta_0$;
\item $M_1$, the mass that the constrained trajectory exceeds at redshift $z_1$, which enters equations (\ref{eq:constrained_p}) and (\ref{eq:constrained_relative_p}) via $S_1$;
\item $z_1$, the redshift at which the constrained trajectory exceeds mass $M_1$, which enters equations (\ref{eq:constrained_p}) and (\ref{eq:constrained_relative_p}) via $\delta_1$.
\end{enumerate}

\subsection{Examples of first-crossing rate distributions}
\label{sec:first-crossing_examples}

\begin{figure}
    \centering
    \includegraphics[width=0.5\textwidth]{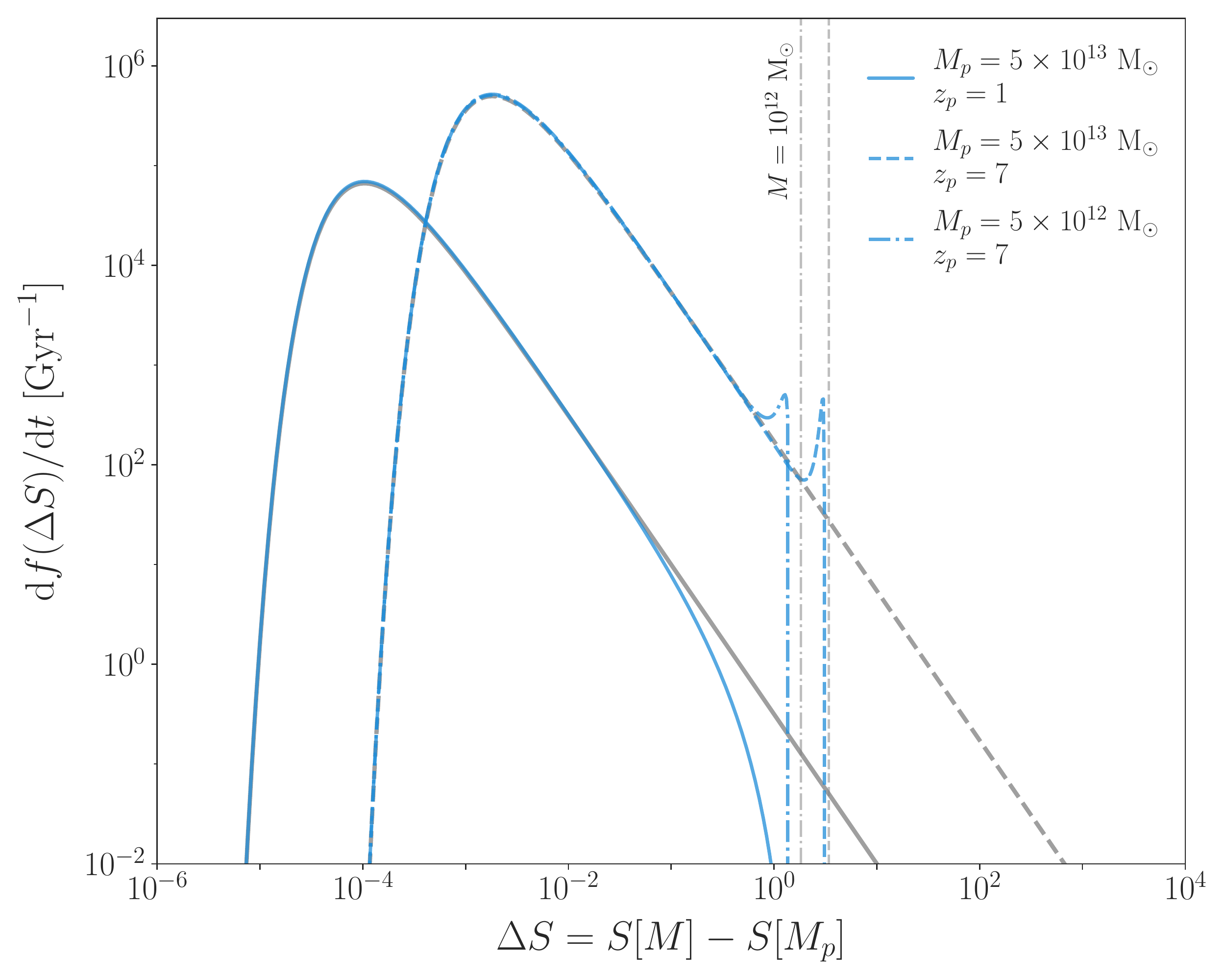}
    \caption{First-crossing rate distributions for constrained Brownian bridge excursions with $M_1=10^{12}~\msun$, $z_1=8$, $M_0=10^{14}~\msun$, and $z_0=0$, versus standard unconstrained ePS trajectories with the same final halo mass and redshift (grey). Solutions are shown for parent haloes with $M_p=5\times 10^{13}~\msun$ and $z_p=1$ (solid), $M_p=5\times 10^{12}~\msun$ and $z_p=7$ (dashed), and $M_p=5\times 10^{13}~\msun$ and $z_p=7$ (dot--dashed); the dot--dashed and dashed unconstrained distributions are identical. Vertical lines show progenitor masses of $10^{12}~\msun$, corresponding to the constrained mass, for $M_p=5\times 10^{13}~\msun$ (dashed) and $M_p=5\times 10^{12}~\msun$ (dot--dashed).}
    \label{fig:first_crossing}
\end{figure}

Fig.~\ref{fig:first_crossing} shows examples of the first-crossing rate distribution, obtained using our numerical solution method described in Appendix~\ref{sec:integral_equations}, for a Brownian bridge excursion with $M_1=10^{12}~\msun$ and $z_1=8$ that reaches $M_0=10^{14}~\msun$ at $z_0=0$ (this corresponds to the blue constraint shown in Fig.~\ref{fig:random_walk}). We plot the first-crossing rate distribution as a function of $\Delta S = S[M]-S[M_\mathrm{p}]$, where $M_\mathrm{p}$ is the mass of the parent halo, i.e., the halo branching into smaller objects at a given redshift as we work backwards in time.

For early times and large parent halo masses, the constrained first-crossing rate is very similar to the unconstrained solution, except that it cuts off at large $\Delta S$. This is expected, because branching events to much smaller masses at early times would violate the Brownian bridge constraint. On the other hand, as $M_\mathrm{p}$ approaches $M_1$ and $z_\mathrm{p}$ approaches $z_1$, the constrained first-crossing rates begin to peak at a $\Delta S$ corresponding to the branching event necessary to satisfy the constraint (formally, in the limit $M_\mathrm{p}\rightarrow M_1$ and $z_\mathrm{p}\rightarrow z_1$, the constrained first-crossing rate approaches a $\delta$ function at the appropriate $\Delta S$). These features lend confidence to our solutions for the constrained first-crossing rate.

\section{Constructing Merger Trees}
\label{sec:merger_tree_construction}
\label{sec:implementation}

We now use the unconstrained and constrained first-crossing rate distributions derived in the previous section to construct merger trees. We describe the standard algorithm for constructing unconstrained merger trees in Section~\ref{sec:standard_algorithm}, our modifications to this algorithm for constructing constrained merger trees in Section~\ref{sec:algorithm_modifications}, and our implementation of the modified algorithm in \textsc{Galacticus} in Section~\ref{sec:galacticus_implementation}. We present examples of constrained and unconstrained merger trees in Section~\ref{sec:merger_tree_examples}.

\subsection{Standard algorithm}
\label{sec:standard_algorithm}

To construct merger trees, we generally follow the procedure described in \cite{Benson12093018}, which is based on the \cite{Parkinson07081382} algorithm. We briefly summarize this procedure below in order to provide context for our constrained merger tree algorithm.

\subsubsection{Branching rates}

In the standard \cite{Parkinson07081382} algorithm, the probability per unit time of a binary
merger occurring for a halo of mass $M$ is
\begin{equation}
    \frac{\mathrm{d}p}{\mathrm{d}\omega} = \int_{M_{\mathrm{res}}}^{M/2} \frac{M}{M'}\frac{\mathrm{d}f[M']}{\mathrm{d}t}\frac{\mathrm{d}S[M']}{\mathrm{d}M'}\left|\frac{\mathrm{d}t}{\mathrm{d}\omega}\right|G[\omega,\sigma(M),\sigma(M')]\mathrm{d}M',\label{eq:df_domega}
\end{equation}
where $\omega\equiv \delta_{c,0}/D(t)$, $M_{\mathrm{res}}$ is the minimum mass resolution of the merger tree, $\mathrm{d}f[M']/\mathrm{d}t$ is the unconstrained first-crossing rate distribution obtained following the procedure in Appendix~\ref{sec:unconstrained_derivation}, and $G[\omega,\sigma(M),\sigma(M')]\equiv G[\omega,\sigma_M,\sigma_{M'}]$ is an empirical modifier function with parameters calibrated to reproduce the progenitor mass function from cosmological simulations. Throughout, we use a generalized form of the \cite{Parkinson07081382} modifier function,
\begin{equation}G[\omega,\sigma_1,\sigma_2]=G_0(\sigma_2/\sigma_1)^{\gamma_1}(\omega/\sigma_1)^{\gamma_2}\left(1 - {\sigma_2^2 \over \sigma_1^2}\right)^{\gamma_3},\label{eq:pch_plus}
\end{equation}
with $G_0=1.14$, $\gamma_1=-0.33$, $\gamma_2=0.06$, and $\gamma_3=0.65$ based on a preliminary recalibration to recent simulations (A.~Benson et al.,\ in preparation). Because we focus on high-mass haloes in this study, the details of the modifier function do not affect our results.

Accretion of haloes below the resolution limit contributes mass to all haloes in the merger tree at a rate
\begin{equation}
    \frac{\mathrm{d}R}{\mathrm{d}\omega} = \int_{M_{\mathrm{min}}}^{M_{\mathrm{res}}} \frac{\mathrm{d}f[M']}{\mathrm{d}t}\frac{\mathrm{d}S[M']}{\mathrm{d}M'}\left|\frac{\mathrm{d}t}{\mathrm{d}\omega}\right|G[\omega,\sigma(M),\sigma(M')]\mathrm{d}M',\label{eq:subresolution}
\end{equation}
where $M_{\mathrm{min}}$ is a minimum mass element, chosen to be sufficiently small such that the resulting progenitor mass distributions converge at the percent level (also see \citealt{Benson12093018}).

\subsubsection{Merger tree construction}

To construct unconstrained merger trees, we select a final halo mass $M_0$ and redshift $z_0$ and we probabilistically generate a series of branching events in which the halo splits into progenitors of lower masses. In particular, when a branching event is determined to occur by comparing a randomly drawn number to the unconstrained branching probability (equation (\ref{eq:df_domega})), the mass of one of the progenitors is drawn from $\mathrm{d}^2f/\mathrm{d}\omega\mathrm{d}M'$ over the interval $[M_{\mathrm{res}},M/2]$, and the remaining progenitor's mass is determined by mass conservation. 

We define the \emph{main branch} as the branch of the merger tree obtained by choosing the more massive progenitor of each merger event (working backwards in time) until the mass resolution limit is reached. In practice, we first construct the main branch and then revisit all secondary progenitors that were generated during construction of the main branch, repeating this procedure to build out the remainder of the unconstrained merger tree.

\subsection{Modifications to the merger tree construction algorithm}
\label{sec:algorithm_modifications}

We modify the \cite{Parkinson07081382} algorithm to build constrained merger trees as follows. 

\subsubsection{Branching rates}

As described by \cite{Cole0007281}, the integral in equation (\ref{eq:df_domega}) spans $[M_{\mathrm{res}},M/2]$ rather than the full mass range $[M_{\mathrm{res}},M]$ of potential progenitor haloes. We Note that, this is an empirical choice, based on which ePS merger trees have been calibrated to cosmological simulations, rather than a requirement of the theory. None the less, it is usually a reasonable approximation because halo merger rates derived from the unconstrained branching rate is almost symmetric about $M/2$ (e.g., see \citealt{Benson0407136,Jiang13115225}). However, our constrained excursions yield branching rates that are generally asymmetric about $M/2$ and can thus break down using the usual~$[M_{\mathrm{res}},M/2]$ interval. For example, as a Brownian bridge excursion approaches the constrained point, the rate approaches a $\delta$-function that guarantees a branching event resulting in the desired progenitor mass $M_1$. Symmetrizing the branching rate by sampling from~$[M_{\mathrm{res}},M/2]$ is therefore not a valid approximation given our constrained first-crossing rate distributions.

To address this, we compute branching rates using the full mass range of $[M_{\mathrm{res}},M-M_{\mathrm{res}}]$ throughout this work,
\begin{equation}
    \frac{\mathrm{d}p}{\mathrm{d}\omega} = \frac{1}{2}\int_{M_{\mathrm{res}}}^{M-M_{\mathrm{res}}} \frac{M}{M'}\frac{\mathrm{d}f[M']}{\mathrm{d}t}\frac{\mathrm{d}S[M']}{\mathrm{d}M'}\left|\frac{\mathrm{d}t}{\mathrm{d}\omega}\right|  \tilde{G}[\omega,\sigma_M,\sigma_{M'}]\mathrm{d}M',\label{eq:df_domega_constrained}
\end{equation}
where the factor of $1/2$ is needed to normalize the branching rate because we draw progenitor masses from the full interval.\footnote{In particular, every branching event adds two progenitors in the \protect\cite{Cole0007281} approach: one with mass $M^\prime$ drawn from the progenitor mass distribution function, and the other with mass $M-M^\prime$ to ensure mass conservation. The overall branching rate must be reduced by a factor 2 to counteract this. In the original \protect\cite{Cole0007281} method, this factor of $1/2$ is implicitly included because the integral spans only half of the full mass range.} In equation (\ref{eq:df_domega_constrained}), $\tilde{G}[\omega,\sigma_M,\sigma_{M'}]$ is our \cite{Parkinson07081382}-like modifier function symmetrized about $M/2$ using the $M<M/2$ half of equation (\ref{eq:pch_plus}), that is
\begin{equation}
\tilde{G}[\omega,\sigma_M,\sigma_{M'}] = \left\{ \begin{array}{ll} G[\omega,\sigma_M,\sigma_{M'}] & \hbox{if } M' \le M/2, \\ G[\omega,\sigma_M,\sigma_{M-M'}] & \hbox{if } M' > M/2.\label{eq:modifier_symmetrized} \end{array} \right.
\end{equation}
We make this choice because the \cite{Parkinson07081382} modifier function (and our version of it in equation (\ref{eq:pch_plus})) is calibrated over the interval $[0,M/2]$ under the assumption of binary mergers.\footnote{As discussed in \cite{Neistein08020198}, binary merger algorithms only approximate the true ePS progenitor mass distribution, which generally includes more than two progenitors per branching event. The binary \cite{Parkinson07081382} algorithm and our modifications thereof have none the less been calibrated to match simulations well, and also predict progenitor mass functions and branching rates that are fairly self-consistent (e.g., see \citealt{Jiang13115225}).} We accordingly modify the sub-resolution accretion rate expression to
\begin{align}
    &\frac{\mathrm{d}R}{\mathrm{d}\omega} = \frac{1}{2} \Bigg(\int_{M_{\mathrm{min}}}^{M_{\mathrm{res}}} \frac{\mathrm{d}f[M']}{\mathrm{d}t}\frac{\mathrm{d}S[M']}{\mathrm{d}M'}\left|\frac{\mathrm{d}t}{\mathrm{d}\omega}\right|\tilde{G}[\omega,\sigma_M,\sigma_{M'}]\mathrm{d}M'&\nonumber \\ & + \int_{M-M_{\mathrm{res}}}^{M-M_{\mathrm{min}}} \frac{\mathrm{d}f[M']}{\mathrm{d}t}\frac{\mathrm{d}S[M']}{\mathrm{d}M'}\left|\frac{\mathrm{d}t}{\mathrm{d}\omega}\right|\tilde{G}[\omega,\sigma_M,\sigma_{M'}]\mathrm{d}M' \Bigg).&\label{eq:subresolution_constrained}
\end{align}

Thus, when a branching event is determined to occur by comparing a randomly drawn number to the branching probability (equation (\ref{eq:df_domega_constrained})), the mass of one of the progenitors is drawn from $\mathrm{d}^2f/\mathrm{d}\omega\mathrm{d}M'$ over the interval $[M_{\mathrm{res}},M-M_{\mathrm{res}}]$, and the remaining progenitor's mass is determined by mass conservation; in the following subsection, we describe how the constrained first-crossing rate distribution is used in these calculations. For consistency, Note that, we also calculate branching rates and sub-resolution accretion using equations (\ref{eq:df_domega_constrained}) and  (\ref{eq:subresolution_constrained}) (rather than equations (\ref{eq:df_domega}) and (\ref{eq:subresolution})) and draw progenitor masses from the full mass interval when constructing unconstrained merger trees (and the unconstrained portions of constrained trees). Given our modifications to the progenitor mass interval, the unconstrained merger trees we construct are not guaranteed to yield halo growth histories that match previous ePS predictions calibrated to cosmological simulations. As discussed below, we defer such recalibration to future work; here, we aim to demonstrate that our constrained merger trees are well converged and satisfy the desired constraints.

\subsubsection{Merger tree construction}

To construct merger trees that satisfy a Brownian bridge constraint, we proceed as follows. We first select the final halo mass, $M_0$, and redshift, $z_0$, as the starting point for our tree construction. We also select the Brownian bridge constrained mass, $M_1$, and redshift,~$z_1$. We then build a specific branch of the tree, referred to as the ``constrained'' branch, by drawing branching events from our constrained first-crossing rate distribution (obtained by plugging equations (\ref{eq:constrained_p}) and (\ref{eq:constrained_relative_p}) into equation (\ref{eq:f_integral_final})). We use the branching and sub-resolution accretion rates in equations (\ref{eq:df_domega_constrained}) and (\ref{eq:subresolution_constrained}), respectively, to build the constrained branch back in time to redshift $z_1$.

\begin{figure}
    \centering
    \includegraphics[width=0.5\textwidth]{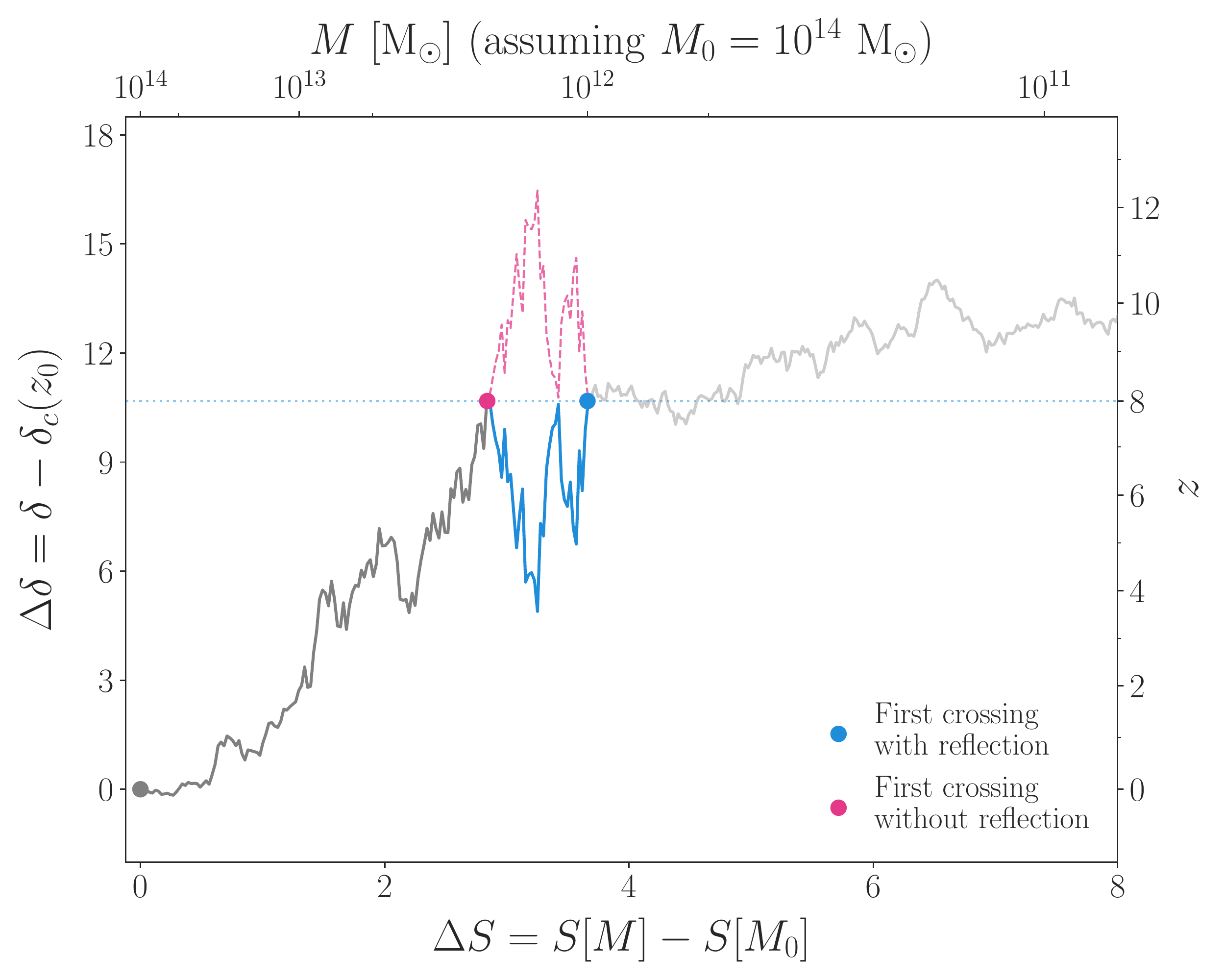}
    \vspace{-4mm}
    \caption{Possible Brownian bridge excursions for a desired constraint of $M_1 = 10^{12}~\msun$ at $z_1=8$ (blue point), for a final halo of mass $M_0=10^{14}~M_{\mathrm{\odot}}$ at redshift $z_0=0$. The dashed pink portion of the excursion first exceeds $\delta_1$ at the pink point, corresponding to a growth history that overshoots the desired constraint. We reflect such excursions about $\delta_1$ to obtain equally probable trajectories, resulting in the solid blue portion of the excursion for the example shown here. This symmetry allows us to insert a merger to a halo of mass exactly $M_1$ at redshift $z_1$ during our constrained merger tree construction. Note that, our constrained merger tree algorithm always uses the reflection principle (corresponding to the blue solid line for the excursion shown here). The bifurcation points are shown for illustrative purposes and do not correspond to physical interactions.}
    \label{fig:random_walk_reflected}
\end{figure}

We then force the constrained branch to split such that one of its progenitors has a mass of \emph{precisely} $M_1$ at redshift $z_1$. This is allowed due to the properties of our constrained Brownian bridge excursions. In particular, these excursions guarantee that $\delta_1$ is first up-crossed at $S\leq S_1$ (for example, see the purple excursion in Fig.~\ref{fig:random_walk} for a case that clearly first up-crosses $\delta_1$ at $S<S_1$). If $\delta_1$ is first up-crossed at $S=S_1$, then a branching event to a progenitor of mass $M_1$ at redshift $z_1$ is directly predicted to occur. Meanwhile, if the constrained excursion reaches $\delta_1$ at $S<S_1$ and then steps towards some $\delta>\delta_1$, this step can be replaced by an equally likely, reflected step towards smaller $\delta$ of equal magnitude, following the classic approach to the ``cloud-in-cloud'' problem in \cite{Bond1991}, based on \cite{Chandrasekhar1943}.\footnote{\cite{Russell13117423} used a similar technique to calculate void statistics in the ``two-barrier'' excursion set formalism, following arguments from \cite{Marchal2003}; also see \cite{Sheth0311260,RussellA13097059}.} Such reflections can be applied for any step that would exceed $\delta_1$ for $S<S_1$, until $S_1$ is reached, guaranteeing a branching event to a progenitor of mass precisely $M_1$ at redshift $z_1$. Thus, we effectively create constrained excursions that \emph{first} up-cross $\delta_1$ at $S_1$.\footnote{Such excursions are formally equivalent to Brownian meanders, i.e., Brownian bridges with a supremum of $\delta_1$ at $S_1$ (also see \citealt{Russell13117423}).} Fig.~\ref{fig:random_walk_reflected} illustrates an example of this reflection principle for possible excursions corresponding to a $M_1=10^{12}~\msun$, $z_1=8$ constraint. 

Two conditions must be met for these reflected steps to be equally likely compared to steps that first up-cross $\delta_1$ for $S<S_1$. First, the barrier must be constant as a function of $\Delta S$, which holds for all scenarios we consider. Second, the Brownian bridge drift term (equation (\ref{eq:relative_drift})) must be zero, which holds because a constrained excursion that has reached $(S,\delta_1)$ at $S<S_1$ is also a Brownian bridge that ends at $(S_1,\delta_1)$, with a new starting point of $(S,\delta_1)$. Thus, the constrained branch is allowed to split to a progenitor of mass \emph{precisely} $M_1$ at redshift $z_1$ (working backwards in time) while maintaining the correct progenitor mass distribution for $S<S_1$. We emphasize that constrained merger trees cannot self-consistently be constructed simply by forcing such branching events to occur using the unconstrained first-crossing rate distribution. In particular, our solution for the constrained first-crossing rate distribution guarantees that the resulting progenitor mass distribution is equivalent to the distribution of unconstrained trajectories down-sampled to satisfy the relevant constraint.

In practice, when a branching event to a progenitor with~$M>M_1$ is drawn along the constrained branch, we manually insert a branching event to a progenitor with a mass of precisely $M_1$ at a redshift of $z_1$. The remainder of the constrained branch (i.e., the~$z>z_1$ portion) is constructed using the unconstrained first-crossing rate distribution. We then return to all secondary haloes and build out the merger tree using the unconstrained first-crossing rate distribution. Thus, except for the constrained branch at redshifts $z<z_1$, all branches of the merger tree are built using the unconstrained first-crossing rate distribution. As in the unconstrained case, all branches (including the constrained branch) are built backwards in time until they reach the mass resolution limit.

Our procedure to stitch together constrained and unconstrained first-crossing rate distributions along the constrained branch, at $z_1$, warrants discussion. Within the scope of the standard uncorrelated random walk excursion set model, this choice implies that trajectories for $S > S_1$ are fairly sampled from all possible trajectories that lead to $(\delta_1,S_1)$. Note that, for $S>S_1$, the statistics of excursions obtained by stitching unconstrained solutions onto our constrained trajectories differ from the distribution of unconstrained excursions only because their ``origin'' is shifted to $(\delta_1,S_1)$; for example, see the trajectories in Fig.~\ref{fig:random_walk}. Thus, unconstrained trajectories that happen to pass through $(\delta_1,S_1)$ are drawn from the same distribution as our constrained trajectories for $S > S_1$. Furthermore, for $S\gg S_1$, our constrained and unconstrained excursions tend towards a mean difference of $\delta_1$ that becomes negligible compared to the variance $S-S_1$, implying that our constrained excursions converge to the unconstrained solutions in this limit. Similarly, for large $S_1$ and sufficiently large $\delta_1$ (meaning that the constrained branch must exceed a small mass at early times), the statistics of our constrained and unconstrained excursions precisely match. We demonstrate that our constrained halo growth histories converge to the unconstrained results in the relevant limits in Appendix~\ref{sec:convergence_tests}.

\begin{figure*}
    \centering
    \includegraphics[width=\textwidth,trim={0 2.5cm 0 1.5cm}]{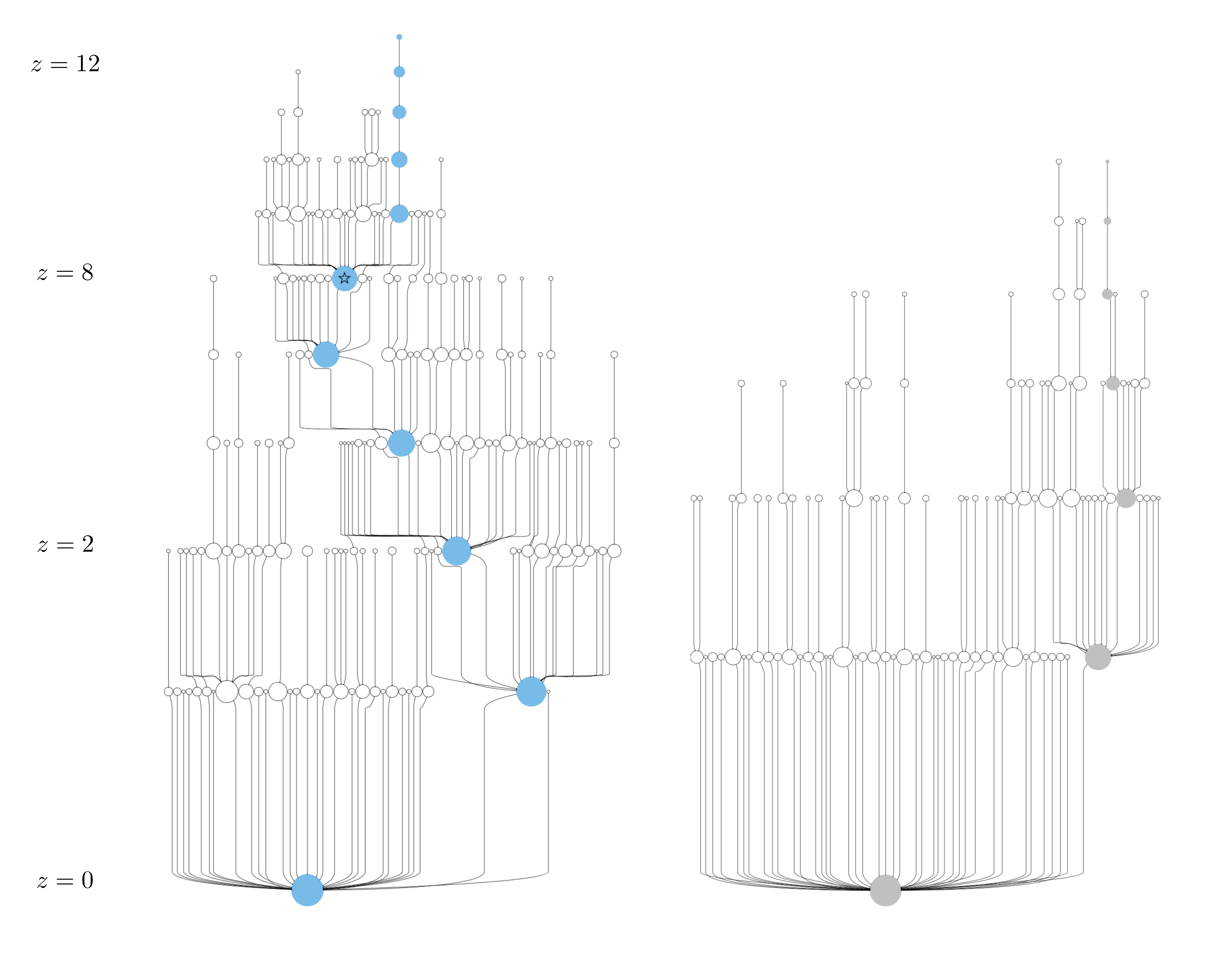}
    \caption{Examples of constrained (left) and unconstrained (right) merger trees for a final halo of mass $M_0=10^{14}~\msun$ at $z_0=0$, which corresponds to the bottom node of each tree. The constrained achieves a mass of precisely $M_1=10^{12}~\msun$ by $z_1=8$; the progenitor at this time is marked by a star. The size of each halo is scaled by the logarithm of its mass, and the edge lengths are proportional to the logarithm of the time elapsed between successive redshifts. The constrained branch of the constrained tree is marked in blue, and the main branch of the unconstrained tree is marked in grey. For illustration, we only show haloes down to a mass of $10^{12}~\msun$; progenitor haloes above this mass in the constrained (unconstrained) tree first appear at $z\approx 12$ ($z\approx 8$).}
    \label{fig:merger_tree}
\end{figure*}

\subsection{Implementation in $\textsc{Galacticus}$}
\label{sec:galacticus_implementation}

We implement the algorithm described above in the open-source semi-analytic model \textsc{Galacticus} \citep{Benson10081786}, and we make our code publicly available at \url{https://github.com/galacticusorg/galacticus}.\footnote{Specifically, we use commit \href{https://github.com/galacticusorg/galacticus/tree/5ea7873f6a3a1ef08f67d114710b77cca7d9529b}{5ea7873f6a3a1ef08f67d114710b77cca7d9529b}.} Here, we briefly describe relevant implementation details, extensions to standard \textsc{Galacticus} excursion set solvers, and numerical choices, and we refer the reader to the extensive online documentation for further details. Note that, all results presented here are from dark matter -- only merger trees; \textsc{Galacticus} semi-analytic galaxy formation models have not been run on top of these merger trees, although the dark matter merger tree structure is unchanged when baryons are added.

In \textsc{Galacticus}, we add functionality that marks branches of constrained merger trees according to whether they were drawn from the constrained branching rate distribution or not. ePS-based merger tree construction algorithms often analyse the ``main'' branch of each merger tree -- namely, the branch defined by selecting the more massive halo progenitor at every branching event, working backwards in time from the final halo at $z_0$. In practice, for constrained solutions with parameters $(M_1,z_1)$, we define the ``constrained'' branch of each merger tree as the branch that directly follows the constrained solution for $S\leq S_1$, joined with the main branch rooted at $(M_1,z_1)$ (the latter portion of each tree is generated using the unconstrained solution for $S>S_1$).

During merger tree construction, we choose sufficiently small time-steps such that neither the maximum probability for a binary merger nor the maximum fractional change in mass due to sub-resolution accretion exceed $10$ per cent at any time. We also employ the time-step optimization algorithm described in Appendix A of \cite{Benson181206026}, which speeds up our merger construction at the $10$ per cent level without loss of accuracy. Throughout, we set $M_{\mathrm{min}}=10^6~\msun$, which is well below any halo mass that directly enters the merger trees we present.

\begin{figure*}
    \centering
    \includegraphics[width=0.475\textwidth]{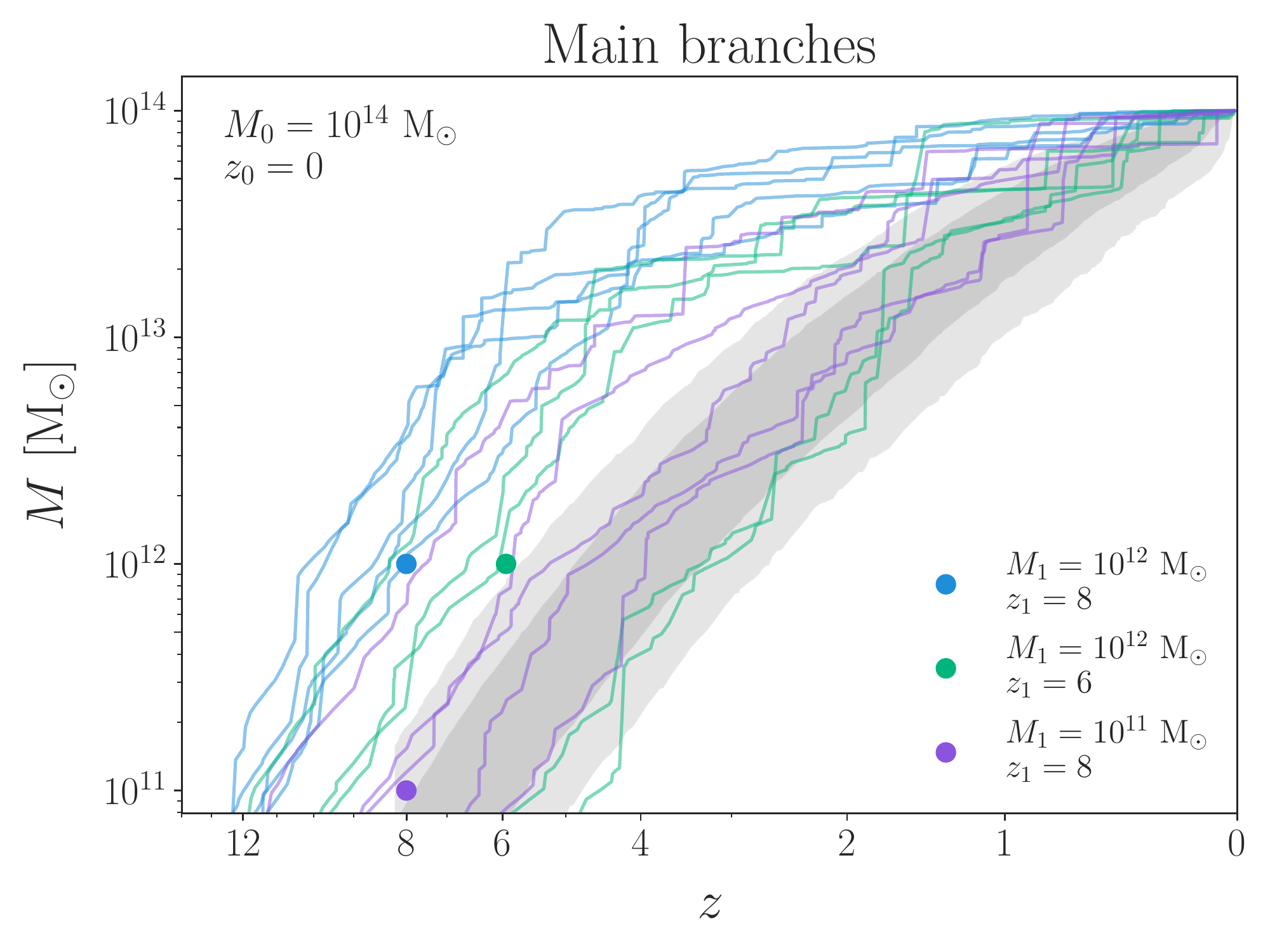}
    \includegraphics[width=0.475\textwidth]{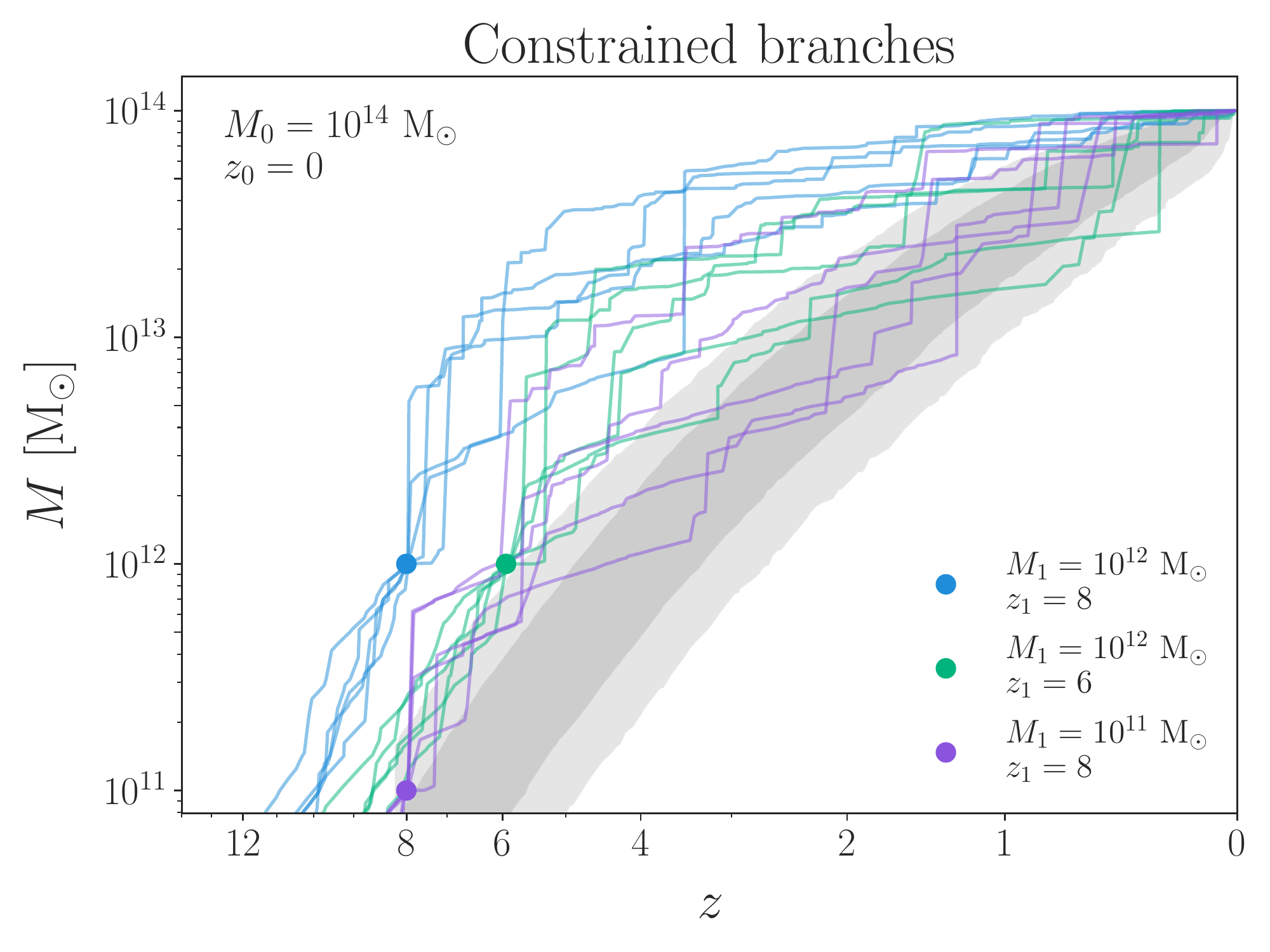}
    \caption{Growth histories for a final halo of mass $M_0=10^{14}~\msun$ at $z_0=0$. Five constrained solutions each are shown for $(M_1,z_1)=(10^{12}~\msun,8)$ (blue), $(10^{12}~\msun,6)$ (green), and $(10^{11}~\msun,8)$ (purple); the dark (light) grey band shows $68$ per cent ($95$ per cent) quantiles of the distribution of unconstrained main branch growth histories. The left-hand panel shows the main branches of the constrained merger trees, and the right-hand panel shows the corresponding constrained branches (see Section~\ref{sec:galacticus_implementation} for definitions). The constrained branches pass precisely through the constrained points; in our implementation, these branches are guaranteed to achieve a mass of precisely $M_1$ at redshift $z_1$. However, Note that, the constrained branch progenitor is not necessarily on the main branch at $z_1$; this implies that main branch progenitors can be less massive than $M_1$ at redshift $z_1$.}
    \label{fig:mah_illustration}
\end{figure*}

\subsection{Examples of constrained merger trees and halo growth histories}
\label{sec:merger_tree_examples}

Fig.\ \ref{fig:merger_tree} shows examples of constrained (left) and unconstrained (right) merger trees for a final halo of mass $M_0=10^{14}~\msun$ at~$z_0=0$. The constrained merger tree is generated using $M_1= 10^{12}~\msun$ and $z_1=8$. This extreme constraint forces the constrained branch to grow to $10^{12}~\msun$ in only $\approx 0.62~\mathrm{Gyr}$, corresponding to the blue point shown in Fig.~\ref{fig:random_walk} (however, the merger tree in Fig.\ \ref{fig:merger_tree} does not correspond to the exact excursion in Fig.~\ref{fig:random_walk}). The constrained branch of this tree is marked in blue, while the main branch of the unconstrained tree is marked in grey. Because we only show progenitors above a mass threshold of $10^{11}~\msun$ at a discrete set of redshifts in Fig.~\ref{fig:merger_tree}, the constrained branch coincides with the main branch of the constrained tree in the visualization. In general, the constrained branch is not guaranteed to merge directly onto the main branch; we discuss this point below.

The growth history of the constrained merger tree in Fig.~\ref{fig:merger_tree} is  clearly very different than a typical unconstrained growth history for this final halo. In particular, the constrained merger tree experiences very little late-time growth, and a significant portion of its final mass is in place by the constrained redshift of $z_1=8$. As a consequence, the progenitors of the constrained final halo above the mass resolution limit of $10^{11}~\msun$ that we use to visualize the merger tree extend significantly farther back in time, to $z\approx 12$, relative to the unconstrained progenitors above this mass that only extend to $z\approx 8$. Visual inspection of many constrained and unconstrained merger trees indicates that this behaviour is typical for the extreme constraint used in Fig.~\ref{fig:merger_tree}. We explore all of the trends discussed above quantitatively in Section~\ref{sec:results}.

Fig.\ \ref{fig:mah_illustration} shows several growth history realizations for a final halo of mass $M_0=10^{14}~\msun$ at $z_0=0$ using the sets of constraint parameters shown in Fig.\ \ref{fig:random_walk} and a mass resolution of~$10^{10}~\msun$. These constrained growth histories exhibit the expected behaviour, in that constraints with larger $M_1$ and/or larger $z_1$ yield growth histories that are more extreme relative to the unconstrained distribution. Decreasing either $M_1$ or $z_1$ tends to push growth histories towards the unconstrained distribution by decreasing the amount of early growth, as quantified in Section~\ref{sec:results}. Note that, the constrained branches pass precisely through their respective $(M_1,z_1)$ constraints, as desired.

We reiterate that the main branches shown in Fig.\ \ref{fig:mah_illustration} do not necessarily exceed a mass of $M_1$ at redshift $z_1$, because (unlike for the constrained merger tree shown in Fig.\ \ref{fig:merger_tree}) each constrained branch does not necessarily merge directly onto its tree's main branches. The choice of a ``preferred'' branch for analysis is less obvious -- and more application-dependent -- for constrained merger trees compared to standard unconstrained trees. In particular, the constraints we consider often increase the rate of early major mergers, for which identifying an unambiguous ``main'' progenitor can be difficult. For example, we could define a branch based on the most massive progenitor of the parent halo at every redshift, regardless of whether this selects a connected sequence of merger events. The quantitative differences between this approach and our calculations based on standard main branches are generally small; the differences are only evident in the rare cases when main branches build up their mass at late times and experience a low-redshift major merger with the constrained branch (e.g., see the latest-forming main branches in the left-hand panel of Fig.~\ref{fig:mah_illustration}). 

Thus, we use main branches when studying the statistical properties of our constrained solutions below. In our high-redshift galaxy application (Section~\ref{sec:applications}), we explicitly track both the constrained and main branches to determine the fates of early galaxies represented by constrained branches in our merger trees.

\section{Results}
\label{sec:results}

\begin{figure*}
    \centering
    \includegraphics[width=0.475\textwidth]{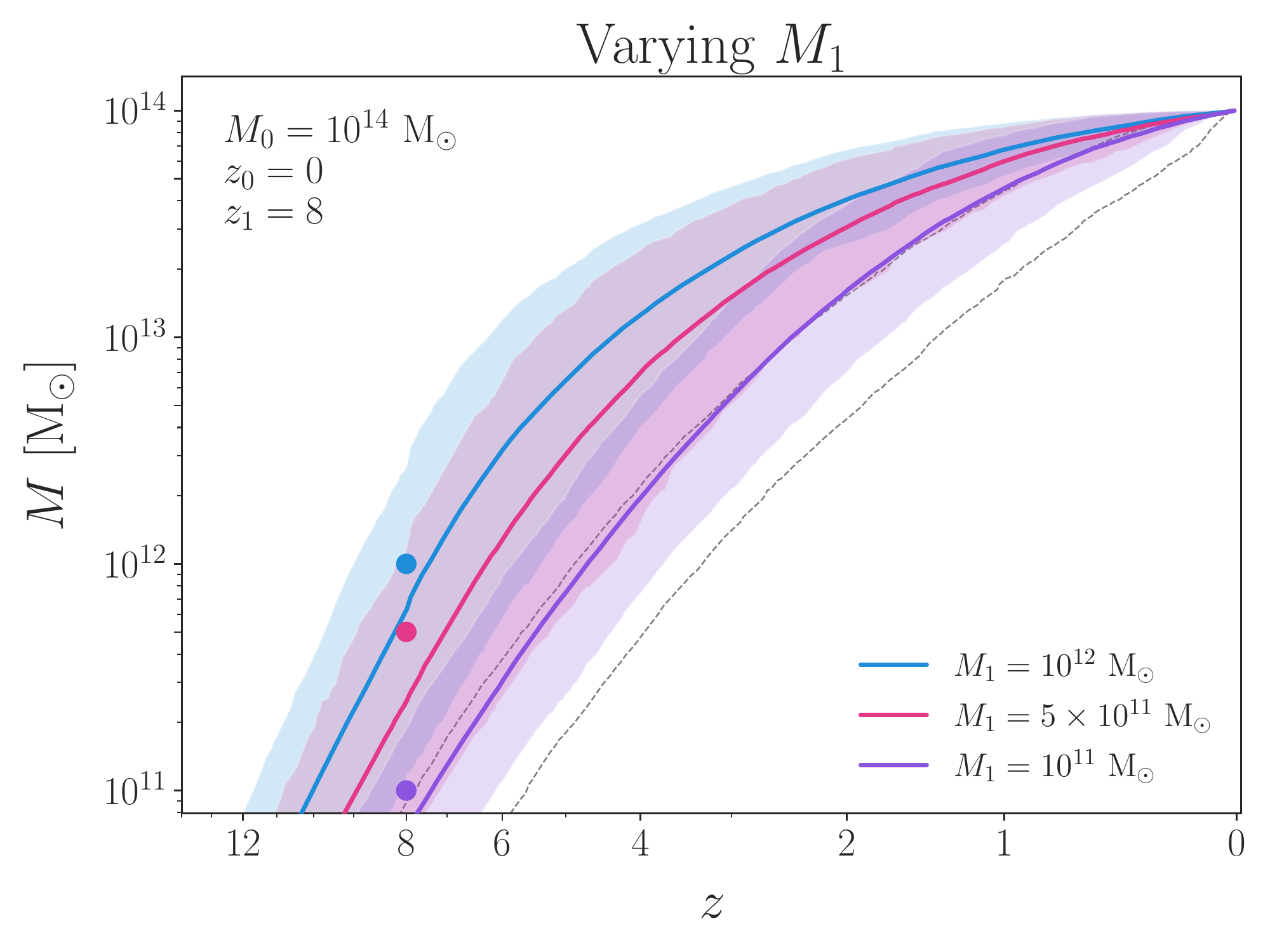}
    \includegraphics[width=0.475\textwidth]{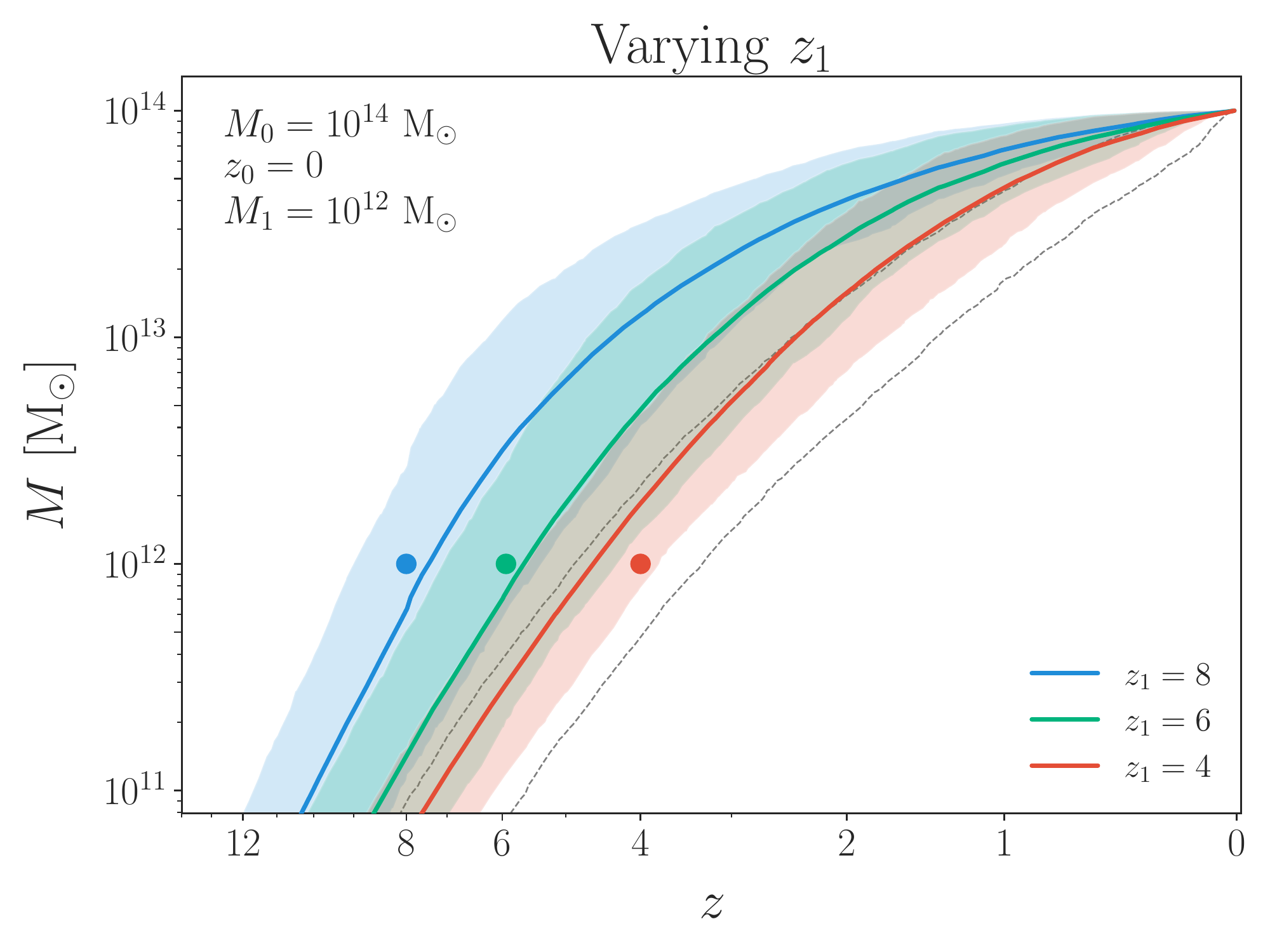}
    \\
    \includegraphics[width=0.475\textwidth]{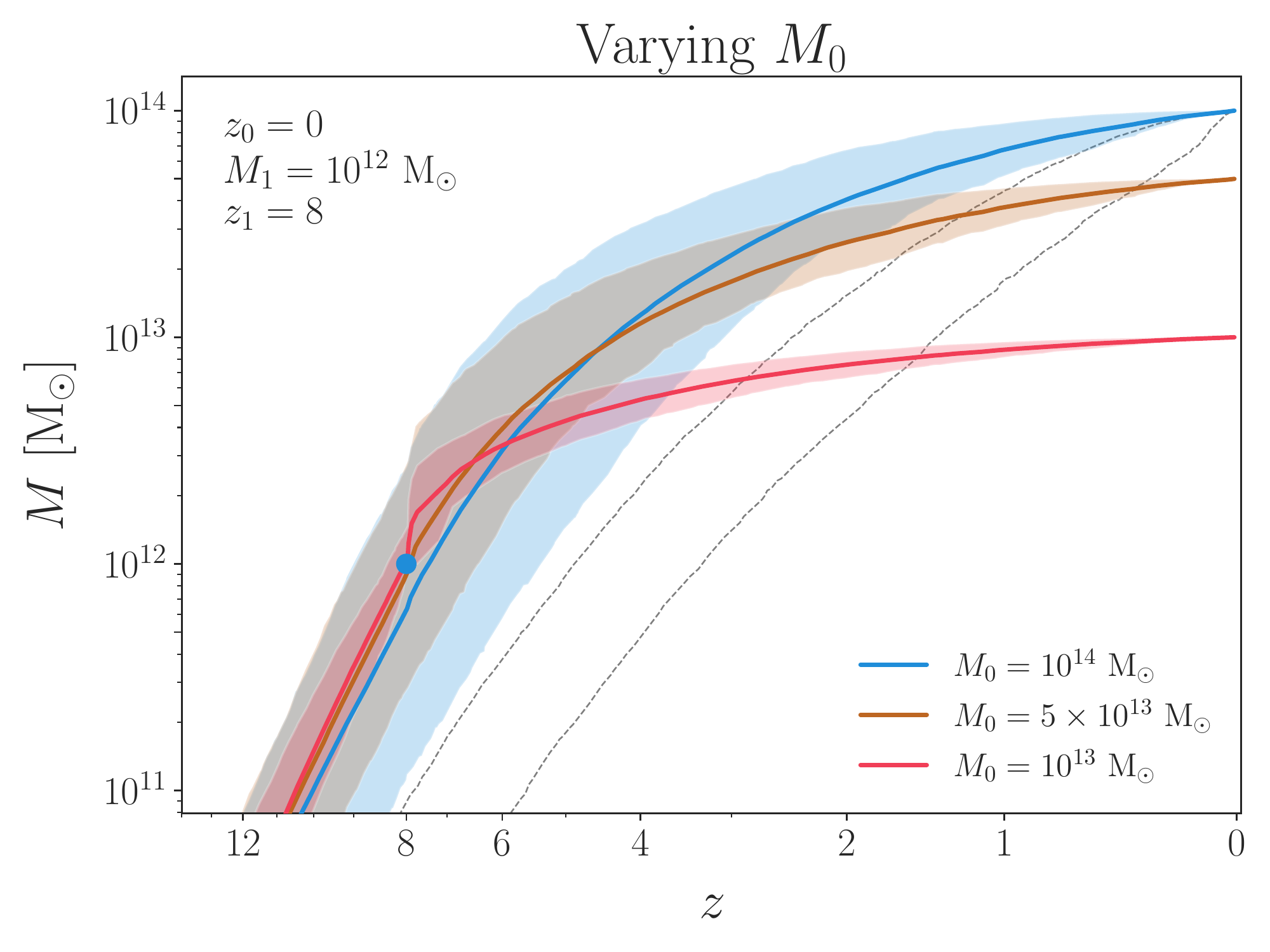}
    \includegraphics[width=0.475\textwidth]{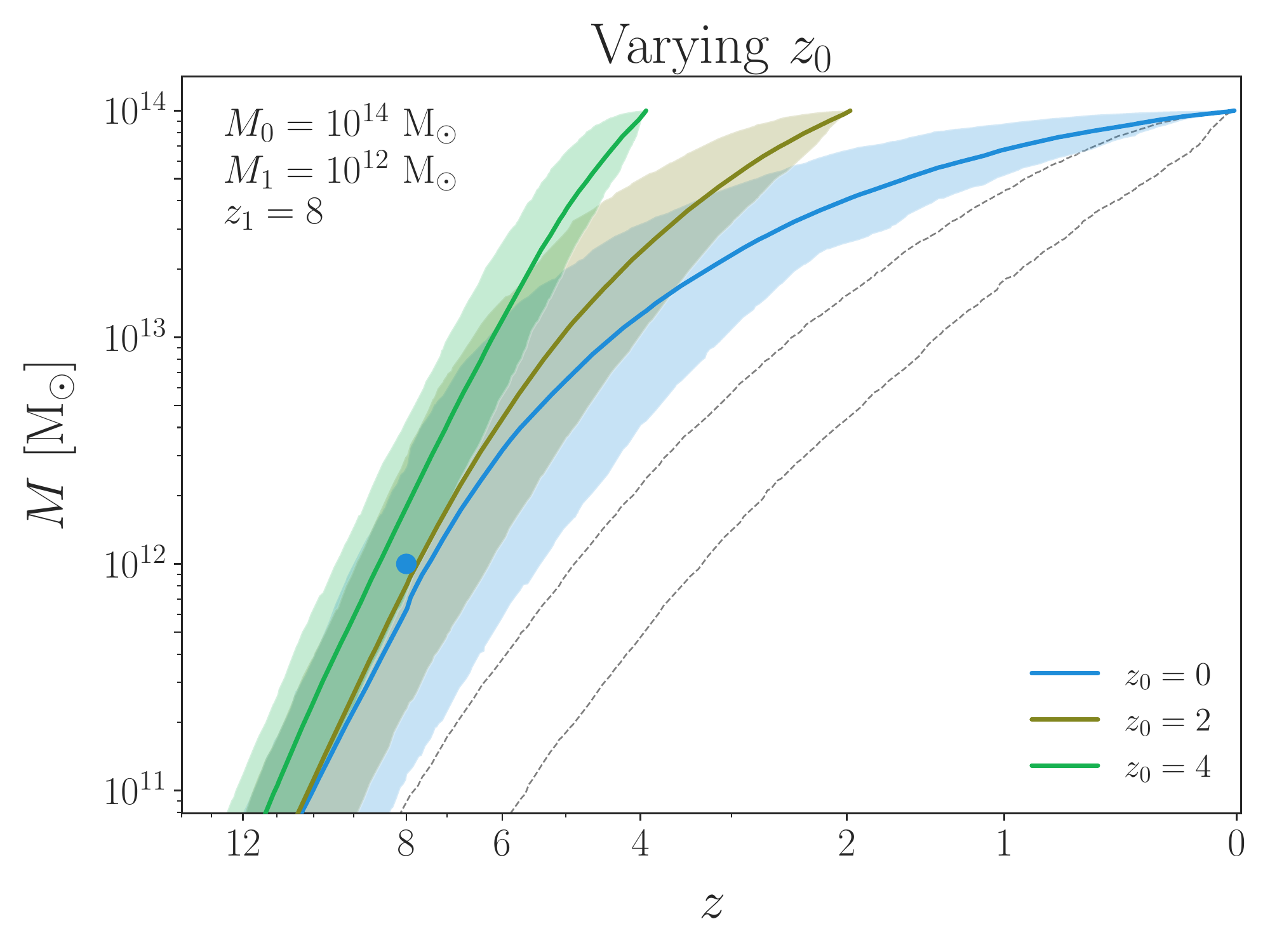}
    \caption{Growth histories for variations in constraint parameters (top panels) or final halo parameters (bottom panels). \emph{Top-left}: Growth histories for final haloes of $M_0=10^{14}~\msun$ at $z_0=0$ whose constrained branches reach a mass of $10^{12}~\msun$ (blue), $5\times 10^{11}~\msun$ (pink), and $10^{11}~\msun$ (purple) at $z_1=8$, listed from earliest to latest-forming. \emph{Top-right}: Growth histories for final haloes of $M_0=10^{14}~\msun$ at $z_0=0$ whose constrained branches reach a mass of $10^{12}~\msun$ at $z_1=8$ (blue), $6$ (green), and $4$ (red). \emph{Bottom-left}: Growth histories for final haloes at $z_0=0$ whose constrained branches reach a mass of $M_1 = 10^{12}~\msun$ at $z_1=8$, for final halo masses of $M_0=10^{14}~\msun$ (blue), $5\times 10^{13}~\msun$ (brown), and $10^{13}~\msun$ (red). \emph{Bottom-right}: Growth histories for final haloes of $M_0=10^{14}~\msun$ whose constrained branches reach a mass of $10^{12}~\msun$ at $z_1=8$, for final halo redshifts of $z_0=4$ (dark green), $2$ (olive), and $0$ (blue). In all panels, the coloured bands shows $68$ per cent quantiles of the growth history distribution for $1000$ constrained realizations, and the dotted grey lines indicate the corresponding quantiles for~$1000$ unconstrained realizations with a final halo mass of $10^{14}~\msun$ at $z_0=0$, for reference.}
    \label{fig:mah_constraints}
\end{figure*}

We now describe our main results, focusing on our predictions for constrained halo growth histories in Section~\ref{sec:growth_histories}, formation times in Section~\ref{sec:formation_times}, and merger rates in Section~\ref{sec:merger_rates}. Throughout this section, we only analyse the main branches of merger trees, allowing for a direct comparison between the constrained and unconstrained cases.

\subsection{Halo growth histories}
\label{sec:growth_histories}

Fig.~\ref{fig:mah_constraints} illustrates the dependence of halo growth histories on the four free parameters chosen to construct constrained solutions: the mass $M_1$ that the constrained branch must reach (top-left-hand panel), the constrained redshift $z_1$ (top-right-hand panel), the mass of the final halo $M_0$ (bottom-left-hand panel), and the redshift of the final halo $z_0$ (bottom-right-hand panel). All panels show~$68$ per cent quantiles of constrained halo growth history distributions, calculated using $1000$ constrained realizations, along with the corresponding quantiles for a fixed set of $1000$ unconstrained realizations that use $M_0=10^{14}~\msun$ and $z_0=0$, for reference. All realizations use a mass resolution of $10^{10}~\msun$ (see Appendix~\ref{sec:convergence_tests} for convergence tests).

Beginning with the upper-left-hand panel of Fig.~\ref{fig:mah_constraints}, we vary $M_1$ for a final halo of mass $M_0=10^{14}~\msun$ and redshift $z_0=0$, at fixed $z_1=8$. We find that increasing $M_1$ shifts halo growth histories systematically earlier, amplifying haloes' phases of rapid early growth and leading to more quiescent behaviour at late times. These qualitative trends are expected based on the requirement to achieve masses of $M_1$ and $M_0$ at specific redshifts, but the detailed shape of the constrained growth histories is a non-trivial prediction of our excursion set solutions. For example, sudden major mergers onto the main branch near $z_1$ that do not shift the entire growth history earlier would also satisfy our constraints. However, such histories are very unlikely compared to ones that feature more overall growth than unconstrained excursions at earlier times, which can be seen from the sample of individual growth histories shown in Fig.~\ref{fig:mah_illustration}. As a result, for a final halo with $M_0=10^{14}~\msun$ at $z_0=0$ and for fixed $z_1=8$, the first haloes along the main branch above $10^{11}~\msun$ (i.e., $10^{-3}$ times the final halo mass) appear at $z\approx 10$, $9$, and $8$ for $M_1=10^{12}$, $5\times 10^{11}$, and $10^{11}~\msun$, respectively, compared to $z\approx 6.5$ in the unconstrained case. We also see that, at fixed $z_1$, decreasing $M_1$ causes the constrained growth histories to approach the unconstrained distribution, as expected. We explicitly demonstrate convergence to the unconstrained solutions in this limit in Appendix~\ref{sec:convergence_tests}.

Next, in the upper-right-hand panel of Fig.~\ref{fig:mah_constraints}, we vary $z_1$ for the same final halo properties, at fixed $M_1=10^{12}~\msun$. Increasing $z_1$ tends to shift growth histories systematically earlier in a coherent, manner similar to the effects of increasing $M_1$. Quantitatively, for a final halo with $M_0=10^{14}~\msun$ at $z_0=0$ and for fixed $M_1=10^{12}~\msun$, the first haloes along the main branch above $10^{11}~\msun$ appear at $z\approx 10$, $9$, and $8$ for $z_1=8$, $6$, and $4$, respectively, again compared to $z\approx 6.5$ in the unconstrained case. At fixed $M_1$, decreasing $z_1$ causes the constrained solutions to approach the unconstrained distribution; again, we demonstrate convergence in this limit in Appendix~\ref{sec:convergence_tests}.

\begin{figure*}
    \centering
    \includegraphics[width=0.475\textwidth]{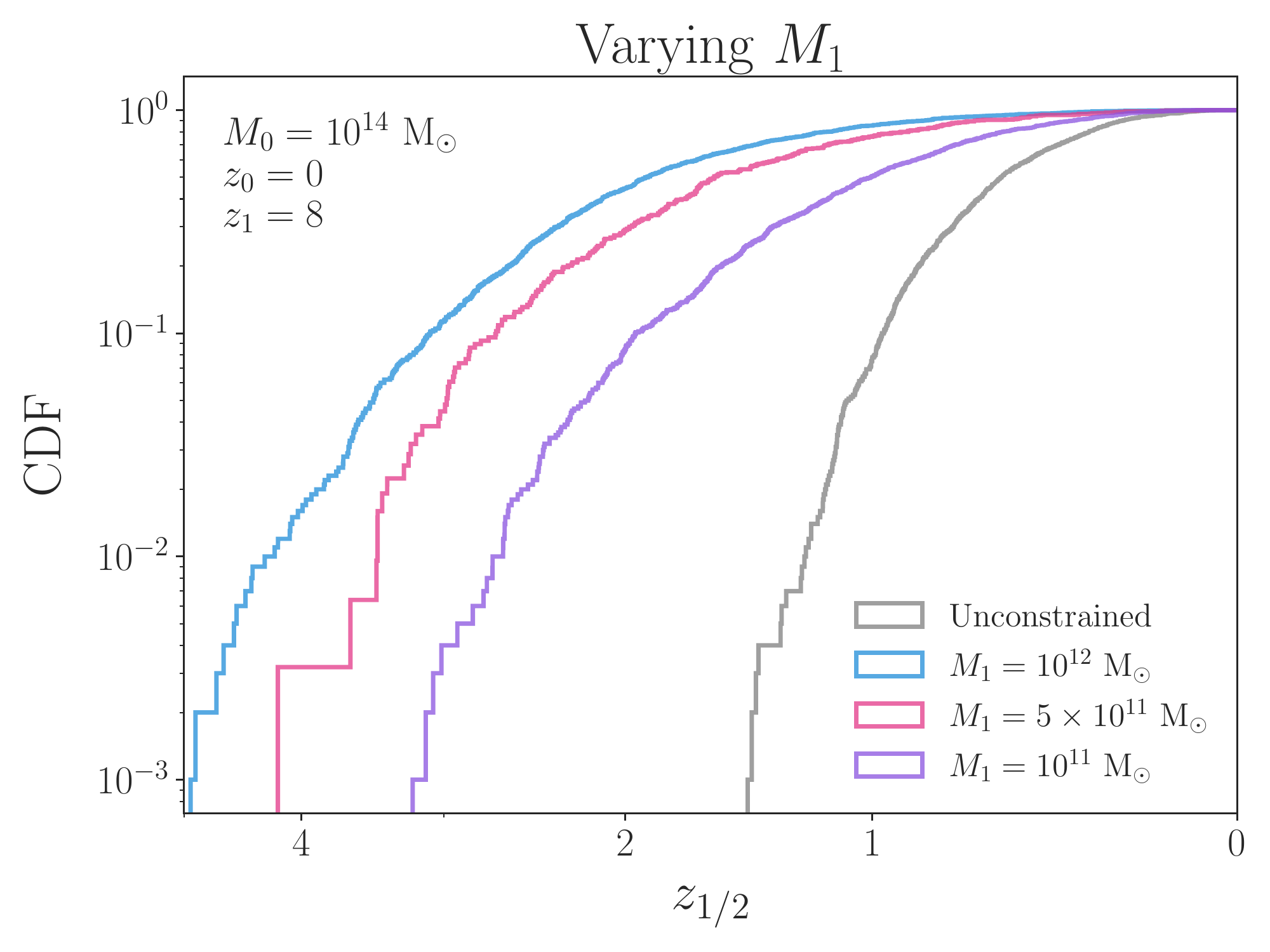}
    \includegraphics[width=0.475\textwidth]{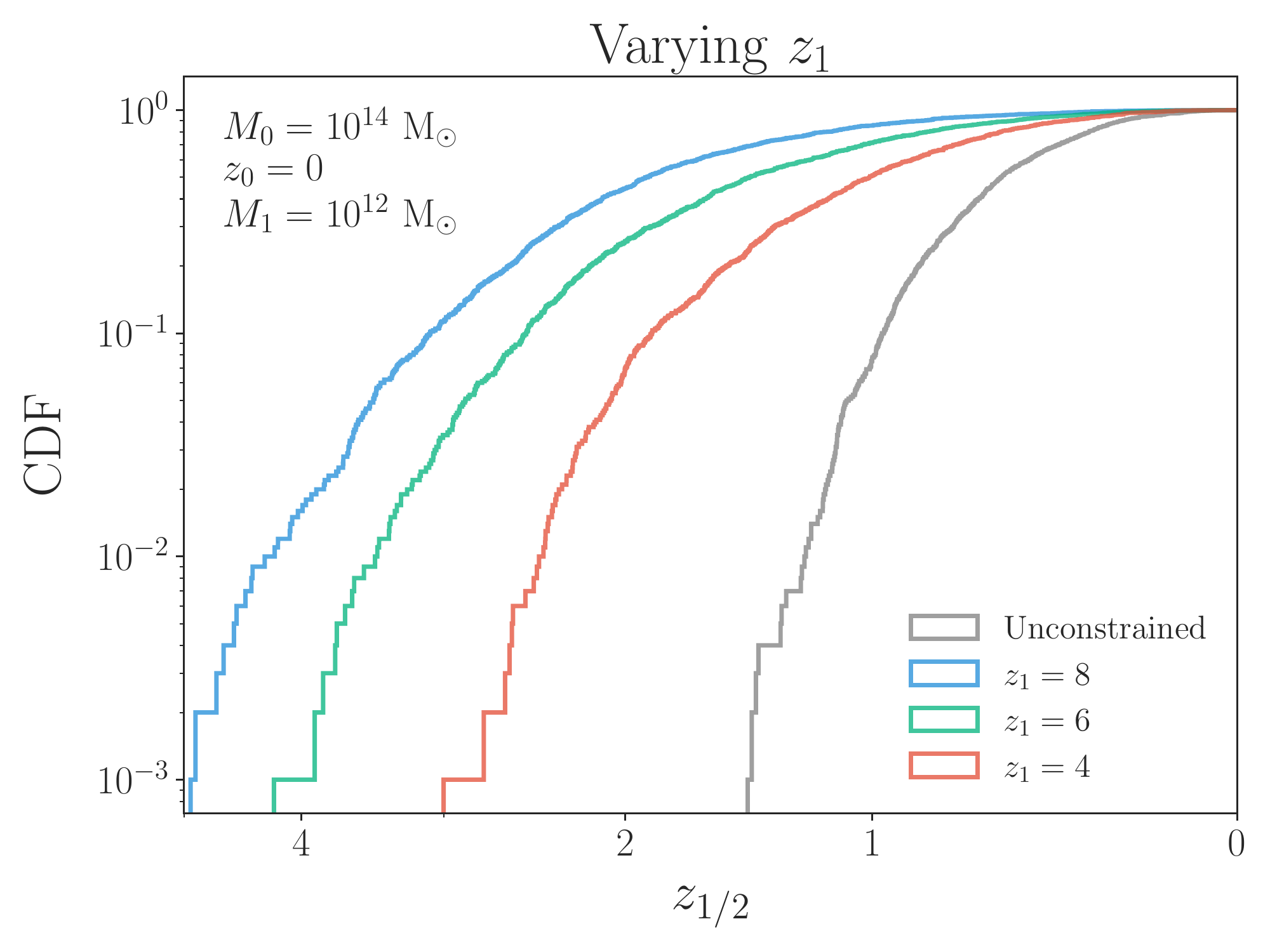}
    \caption{Cumulative distributions of formation time, defined as the redshift at which a halo reaches half of its $z=0$ mass, for final haloes of mass $M_0=10^{14}~\msun$ at redshift $z_0=0$. Constrained realizations for various Brownian bridge parameters are shown by coloured lines, and the cumulative distribution for standard unconstrained excursions is shown by the grey line; each distribution is computed using $1000$ constrained or unconstrained realizations. \emph{Left}: Half-mass formation redshift distributions for constrained merger trees whose constrained branches reach a mass of $10^{12}~\msun$ (blue), $5\times 10^{11}~\msun$ (pink), and $10^{11}~\msun$ (purple) at $z_1=8$, listed from earliest to latest-forming. \emph{Right}: Same as the left-hand panel, for constrained merger trees whose constrained branches reach a mass of $10^{12}~\msun$ at $z_1=8$ (blue), $6$ (green), and $4$ (red).}
    \label{fig:formation_time}
\end{figure*}

For variations in both $M_1$ and $z_1$, we find that a quantitative measure of extremity based on the gradient of the Brownian bridge drift term in equation (\ref{eq:drift_brownian_bridge}) predicts the magnitude of the resulting shifts in growth histories towards earlier or later times fairly well. In particular, the quantity $\mathcal{E}\equiv (\delta_1-\delta_0)/(S_1-S_0)$ correlates with the mean temporal offset of the resulting constrained growth histories relative to the unconstrained distribution.

Finally, the bottom-left and bottom-right-hand panels of Fig.~\ref{fig:mah_constraints} fix constraint parameters of $M_1=10^{12}~\msun$ and $z_1=8$, and vary $M_0$ and $z_0$, respectively. Variations in $M_0$ do not significantly affect the early growth, and instead simply cause the histories to flatten out (i.e., stop growing) at different redshifts. This follows because changing $M_0$ shifts the origin of our constrained excursions but does not affect the statistical properties of the random walks for variances larger than $S_1$ (corresponding to the unconstrained portion of our constrained excursions, as shown in e.g.\ Fig.~\ref{fig:random_walk}). Meanwhile, increasing $z_0$ simply cuts off the growth histories at $z>0$. As expected, we find that constrained growth histories at fixed $M_1$ and~$z_1$ are greater outliers relative to the corresponding unconstrained distributions as $M_0$ is decreased (and approaches $M_1$), and as $z_0$ is increased (and approaches $z_1$).

We note that, based on the halo mass function predicted by our unconstrained ePS calculations, haloes that satisfy our high-redshift constraints are expected to descend to haloes with a specific mass distribution at $z=z_0$; these descendant mass distributions are derived in Appendix~\ref{sec:descendant_derivation}. Choosing values of $M_0$ well outside the expected range of descendant masses alters the resulting growth histories relative to the unconstrained distribution, effectively by making the constraint more extreme. However, even when comparing constrained and unconstrained solutions that descend to a likely halo mass for a given high-redshift constraint, the constrained growth histories statistically differ from their unconstrained counterparts. For example, this is the case for our fiducial choice of $M_0=10^{14}~\msun$ and $z_0=0$ given $M_1=10^{12}~\msun$ and $z_1=8$, and is reflected by a change in the overall morphology of constrained versus unconstrained merger trees, where the constrained cases form earlier (e.g., see Fig.~\ref{fig:merger_tree}).

\subsection{Halo formation times}
\label{sec:formation_times}

 Fig.\ \ref{fig:formation_time} shows distributions of half-mass formation redshift, defined as the redshift, $z_{1/2}$, at which $M(z=z_{1/2})/M(z=0)=0.5$, for final haloes of mass $M_0=10^{14}~\msun$ at $z_0=0$. In particular, we compare~$z_{1/2}$ distributions from the sample of $1000$ unconstrained merger trees described above to constrained predictions with the same $M_0$ and $z_0$, with $1000$ realizations each for several values of $M_1$ (left-hand panel) and $z_1$ (right-hand panel). The unconstrained $z_{1/2}$ distributions have a mean and standard deviation of $0.61\pm 0.27$ and are in reasonably good agreement with predictions from cosmological simulations (e.g., \citealt{Wechsler0108151,Nadler220902675}).

The constrained $z_{1/2}$ distributions behave as expected based on the discussion in the previous subsection. In particular, at fixed $z_1=8$, these distributions have a mean and standard deviation of $1.9\pm 0.9$, $1.6\pm 0.8$, and $1.1\pm 0.6$ for $M_1=10^{12}$, $5\times 10^{11}$, and $10^{11}~\msun$, respectively. These shifts towards earlier formation times result from more rapid early accretion, which is necessary to achieve the constrained mass. Quantitatively, these constrained realizations are thus $\approx 4\sigma$, $3\sigma$, and $2\sigma$ outliers relative to the unconstrained formation time distribution, respectively. 

Meanwhile, at fixed $M_1=10^{12}~\msun$, the constrained formation time distributions have a mean and standard deviation of $1.9\pm 0.9$, $1.5\pm 0.8$, and $1.1\pm 0.5$ for $z_1=8$, $6$, and $4$, again making them $\approx 4\sigma$, $\approx 3\sigma$, and $2\sigma$ outliers in the mean relative to the unconstrained formation time distribution, respectively. Moreover, our constrained solutions produce individual merger trees with formation times that are extreme outliers relative to the unconstrained distribution; for example, the realizations in our $z_1=8$, $M_1=10^{12}~\msun$ case with~$z_{1/2}=4$ that occur about one in a hundred times arise by chance in the corresponding unconstrained trees roughly one in ten billion times, representing a $\sim 10^8$ times speed up.

\subsection{Halo merger rates}
\label{sec:merger_rates}

Having shown that halo growth histories are shifted systematically earlier in the presence of constraints that guarantee the existence of a massive progenitor at early times, we now investigate the distributions of mergers that build up the final haloes in constrained scenarios. In particular, we will demonstrate that high-redshift constraints lead to systematically earlier merging activity and an enhanced rate of minor mergers after the constraint is achieved.

Fig.~\ref{fig:deltaM_z} shows the distribution of merger events along the main branch as a function of redshift, calculated as $\Delta M/M$, where $M$ is the mass of the main branch progenitor at a given redshift and $\Delta M$ is the amount by which $M$ increases during a given merger. For clarity, we only show mergers that occur while the main branch has achieved at least $10$ per cent of its final mass, corresponding to a mass of $10^{13}~\msun$ for the $M_0=10^{14}~\msun$, $z_0=0$ haloes used to generate the results shown in Fig.~\ref{fig:deltaM_z}. Because the resolution of these merger trees is $M_{\mathrm{res}}=10^{10}~\msun$, we can only measure merger events with $\Delta M/M>10^{-4}$ for the solutions discussed here.

Fig.~\ref{fig:deltaM_z} shows all individual mergers from $1000$ realizations of constrained merger trees for $M_1=10^{12}$ and $10^{11}~\msun$, both using $z_1=8$, and we compare these to the underlying distribution for the same number of realizations of unconstrained merger trees. We find that merger times are shifted systematically earlier as $M_1$ increases; even for $M_1=10^{11}~\msun$, for which the constrained and unconstrained halo growth history $68$ per cent quantiles partially overlap (see the top-left-hand panel of Fig.~\ref{fig:mah_constraints}), this shift is significant. For example, the mean and standard deviation of the redshift at which any merger with $\Delta M/M>10^{-2}$ occurs is $z=1.1\pm 0.6$, $z=1.7\pm 0.9$, and $z=2.8\pm 1.7$ for the unconstrained, $M_1=10^{11}~\msun$, and $M_1=10^{12}~\msun$ case, respectively. These shifts are driven by the requirement that the constrained branch progenitor achieves an unusually large mass at early times and then (eventually) merges onto the main branch. To guarantee conservation of mass, there are then fewer late-time mergers in the constrained trees, and the late-time mergers that do occur are biased towards lower mass ratios relative to the unconstrained distribution.

We find that there is a slightly enhanced tail of small mergers in the constrained cases that compensates for major merger events with the constrained branches at early times. None the less, the overall mass distribution of mergers along the main branch does not significantly differ among the cases shown in Fig.~\ref{fig:deltaM_z}, despite their different distributions of merger redshifts. The overall similarity in merger mass distributions is reasonable because most merger events are minor, and the distribution of first-upcrossings of the ePS barrier after shifting the origin of our excursions appropriately, for each constrained case, is not significantly influenced by the constraints.

\section{The Descendants of the First Galaxies}
\label{sec:applications}

We now illustrate their utility in a specific astrophysical context. In particular, we consider the extremely luminous high-redshift galaxy candidates detected by JWST. Because of observational selection effects, these systems represent the most luminous and massive objects in existence at early times, and are therefore outliers relative to the underlying high-redshift galaxy and halo populations. In Section~\ref{sec:highz_halo_mass}, we estimate the halo masses of these high-redshift systems; we interpret these masses, and the corresponding redshifts, as the $M_1$ and $z_1$ parameters entering our Brownian bridge excursions. In Section~\ref{sec:highz_constrained}, we then use our constrained merger tree algorithm to generate the most likely distributions of growth histories for these high-redshift systems. Specifically, we calculate the distribution of $z_0=0$ descendant halo masses given each JWST galaxy candidate's redshift and inferred halo mass, and we draw from these distributions to set $M_0$ in our constrained merger tree realizations.

\subsection{Estimating high-$z$ JWST galaxy candidates' halo masses}
\label{sec:highz_halo_mass}

We consider the seven high-redshift JWST galaxy candidates reported by \cite{Labbe220712446}, which were selected based on the presence of Balmer breaks. We emphasize that these objects' redshifts are not spectroscopically confirmed, and that several observational and modelling systematics may impact the interpretation of their stellar masses and redshifts; thus, we refer to them as galaxy \emph{candidates}.\footnote{As this paper was being prepared, a distinct sample of JWST galaxy candidates were spectroscopically confirmed \citep{Robertson221204480,Curtis-Lake221204568}; furthermore, \cite{Bouwens221206683} reported that the \cite{Labbe220712446} candidates' photometrically derived properties are relatively robust based on consistency between independent analyses of these systems}. Note that, we do not include the even more extreme candidate reported by \cite{Naidu220709434} in our calculations because its high-redshift solution is under scrutiny (e.g., \citealt{Naidu220802794}).

\begin{figure}
    \centering
    \includegraphics[width=0.475\textwidth]{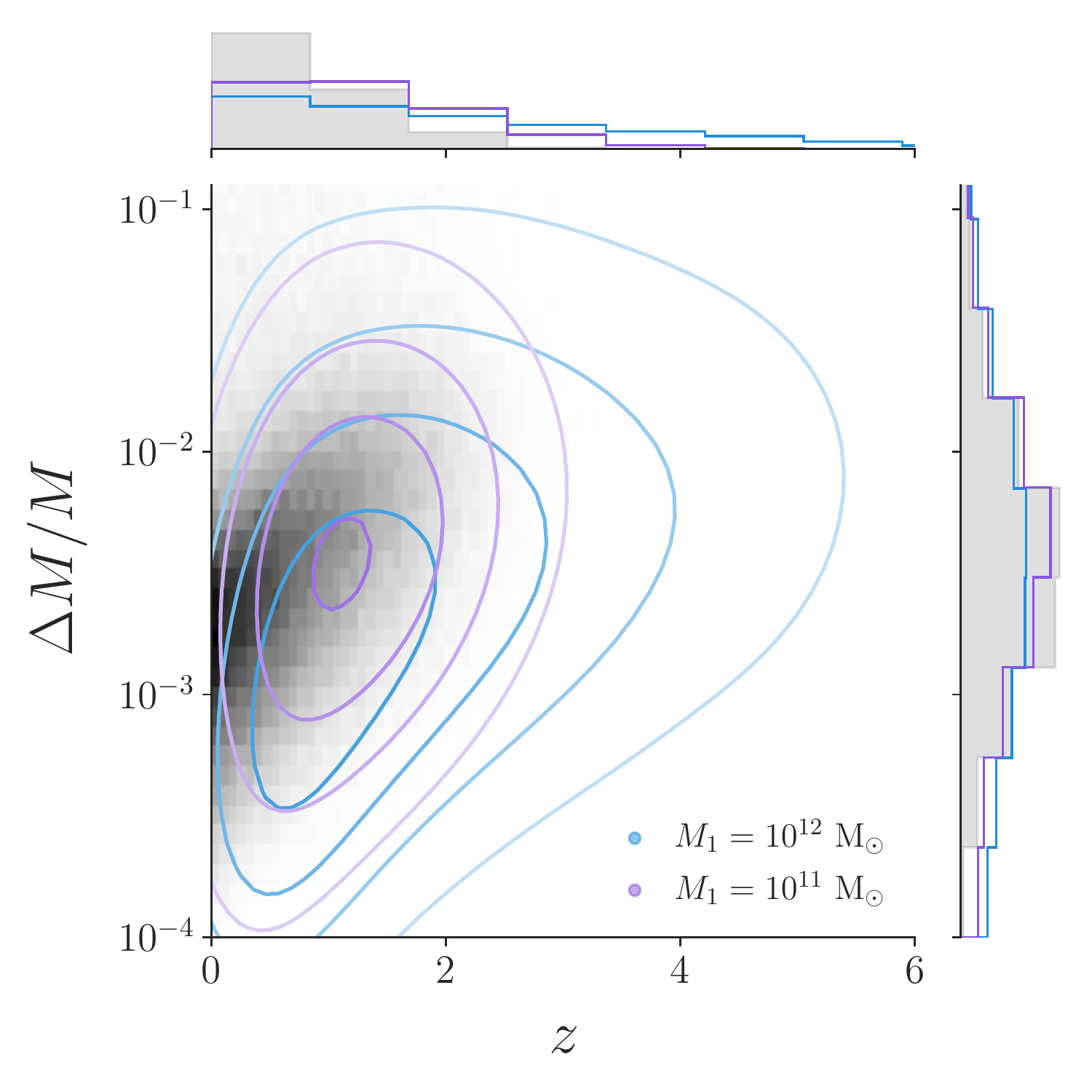}
    \caption{The mass and redshift distribution of mergers along the main branch for $1000$ realizations of constrained merger trees with $M_1=10^{12}~\msun$ (blue) and $10^{11}~\msun$ (purple), both with $z_1=8$, for a final halo of mass $M_0=10^{14}~\msun$ at redshift $z_0=0$. Lines show four iso-proportion contours of the joint probability density, and the corresponding unconstrained distribution is shown as a grey histogram; the top and right-hand panels show the corresponding marginal distributions, where the spread encompasses the four iso-proportion contours. Only mergers that take place when the main branch has achieved at least $10$ per cent of its final mass (i.e. $M>10^{13}~\msun$) are considered; there are typically a few thousand such mergers above our resolution limit of $\Delta M/M>10^{-4}$.}
    \label{fig:deltaM_z}
\end{figure}

\begin{figure*}
    \centering
    \includegraphics[width=0.475\textwidth]{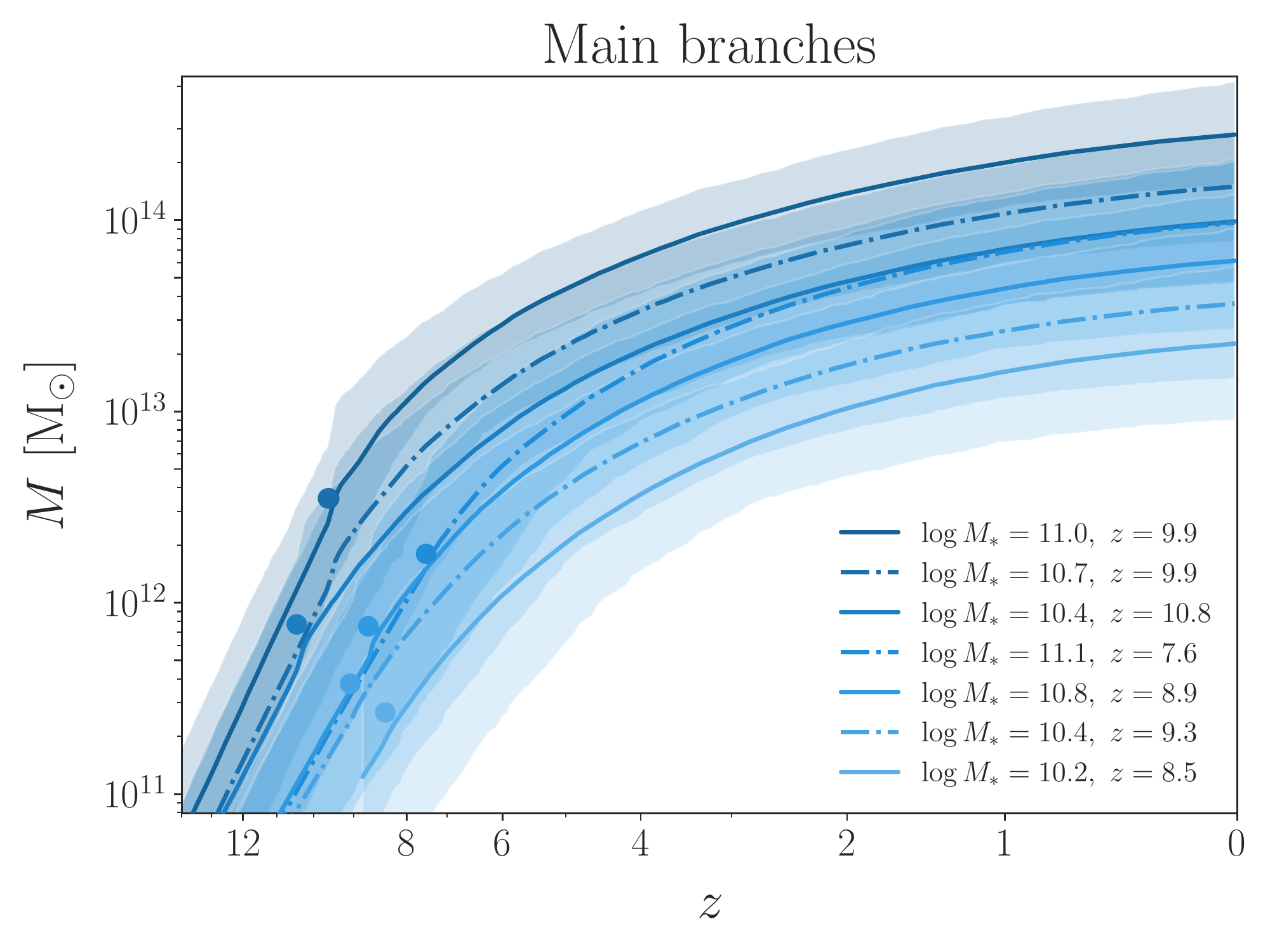}
    \includegraphics[width=0.475\textwidth]{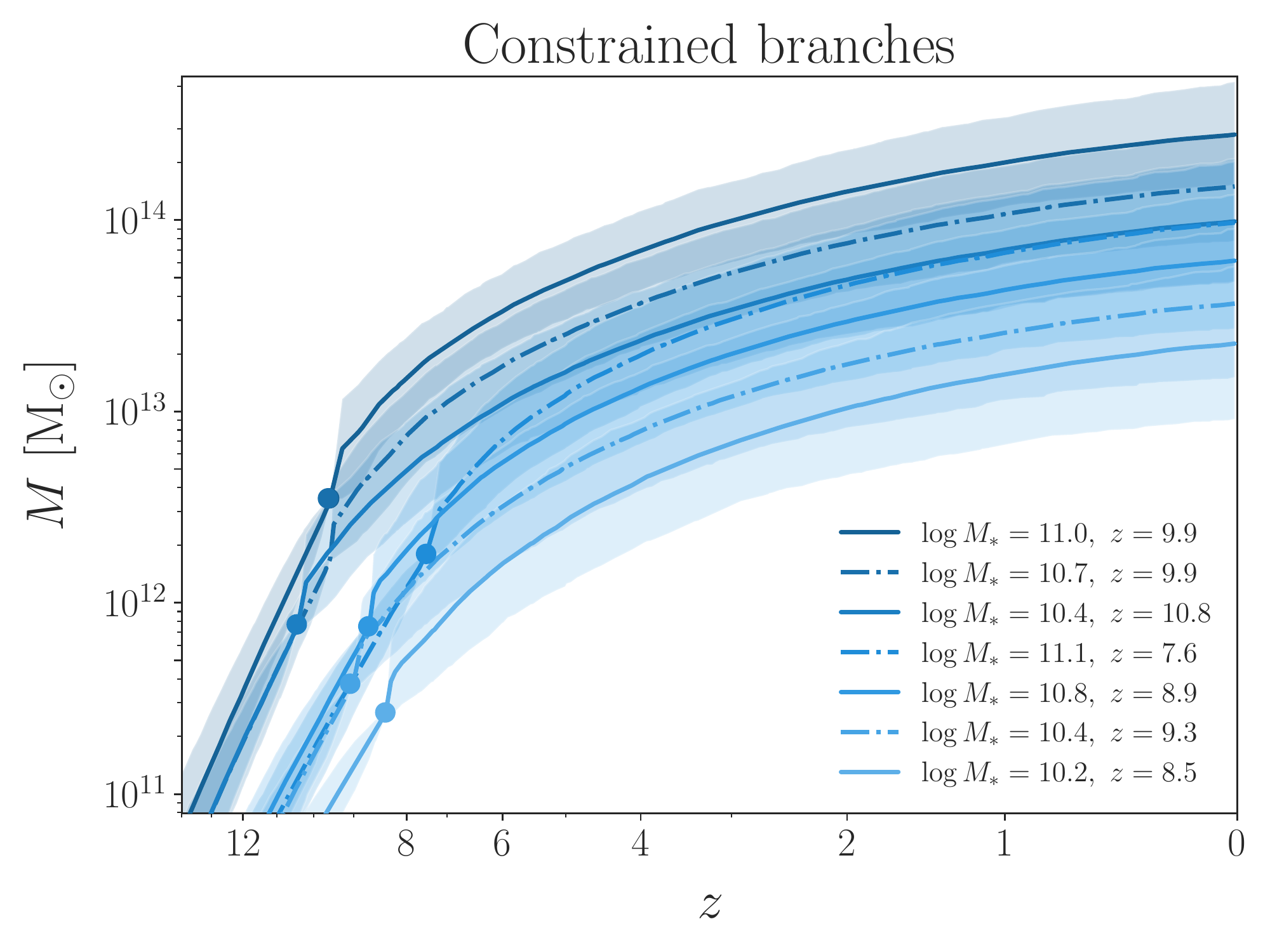}
    \caption{Growth histories for the seven high-redshift JWST galaxy candidates from \protect\cite{Labbe220712446}. Bands shows $68$ per cent quantiles of the distribution of unconstrained main and constrained branch growth histories. The left-hand panel shows the main branches of the constrained merger trees, and the right-hand panel shows the constrained branches (see Section~\ref{sec:galacticus_implementation}). Each constrained halo mass is derived from the corresponding galaxy's estimated stellar mass and photometric redshift, as described in Section~\ref{sec:highz_halo_mass}, and the constrained redshift is taken to be equal to the photometric redshift.}
    \label{fig:jwst_results}
\end{figure*}

Our procedure for estimating halo masses is based on the extreme value statistics (EVS) calculations and code from \cite{Lovell220810479}.\footnote{See \url{https://github.com/christopherlovell/evstats}.} In particular, these authors present distributions of the maximum halo and stellar masses predicted to exist in a fiducial $\Lambda$CDM cosmology as a function of redshift. We calculate the quantile of the stellar mass EVS distribution, at the appropriate redshift, corresponding to each of the \cite{Labbe220712446} galaxy candidates; these systems' reported redshifts and stellar masses span $7.5\lesssim z\lesssim 11$ and $10^{10}~\msun \lesssim M_*\lesssim 10^{11}~\msun$, respectively. We confirm that the quantile of the least (most) extreme galaxy candidate relative to the stellar mass EVS distribution corresponds to a~$\approx 2\sigma$ ($4\sigma$) outlier assuming Gaussian statistics, as shown by \cite{Lovell220810479}.

We then translate these statistics to halo masses by finding the halo mass that minimizes the difference between each galaxy's stellar mass and halo mass EVS quantile, at the appropriate redshift. This yields a distribution of halo masses that spans $2.6\times 10^{11}~\msun\lesssim M_{\mathrm{halo}}\lesssim 3.5\times 10^{12}~\msun$.\footnote{The halo mass function used to calculate the EVS distribution uses the $M_{200}$ mass definition \citep{Lovell220810479}; we do not attempt to convert these masses to our ePS halo mass definition because this $\mathcal{O}(10\ \mathrm{per\ cent})$ difference is small compared to other uncertainties in our estimate.} Specifically, for the seven \cite{Labbe220712446} galaxy candidates with estimated redshifts and stellar masses of $(z_{*},\log(M_{*}/\msun))=(9.9,11.0)$, $(9.9,10.7)$, $(10.8,10.4)$, $(7.6,11.1)$, $(8.9,10.8)$, $(9.3,10.4)$, $(8.5,10.2)$, we estimate halo masses of $\log(M_{\mathrm{halo}}(z_{*})/\msun)=12.5$, $12.2$, $11.9$, $12.2$, $11.9$, $11.6$, $11.4$, respectively. We always order these galaxies in decreasing order of their average expected $z_0=0$ halo masses derived from our calculation in Appendix~\ref{sec:descendant_derivation}. In the following subsection, we use each galaxy candidate's redshift and estimated halo mass as constraint parameters to generate a family of merger trees for the \cite{Labbe220712446} sample.

Before moving on, we discuss the basic assumption underlying our halo mass estimates and areas for improvement. In essence, we have assumed that the haloes of JWST galaxy candidates are equal outliers in halo mass, relative to the underlying high-redshift halo population, as the galaxies themselves are in \emph{stellar} mass, relative to the underlying high-redshift galaxy population. Although this assumption is plausible, it is almost certainly incorrect in detail. Empirically, high-redshift galaxy and halo populations are related via a non-linear galaxy--halo connection with potentially significant scatter, which implies that the one-to-one correspondence our calculation assumes likely breaks down (e.g., see \citealt{Wechsler180403097} for a review). More sophisticated abundance-matching techniques and empirical modelling based on halo populations in cosmological simulations (e.g., \citealt{Behroozi14045299,Behroozi160904402,Behroozi200704988}), or direct estimates of halo properties from hydrodynamic simulations (e.g., \citealt{Lovell220810479}), would thus improve the accuracy of our estimates. Physically, correlations between secondary halo properties (e.g., accretion rate) and galaxy properties that correlate with observable quantities (e.g., star formation rate) imply that a one-dimensional correspondence based on halo and stellar mass is likely inadequate for explaining why some high-redshift haloes host extremely luminous galaxies (e.g., \citealt{Mason220714008,Mirocha220812826}). We therefore proceed with the caveat that our halo mass estimates are subject to improvement. The utility of our constrained merger tree realizations for predicting the growth histories of high-redshift systems is independent of these uncertainties.

\subsection{Constrained excursion set predictions}
\label{sec:highz_constrained}

We integrate our halo mass estimates for the high-redshift JWST galaxy candidates into constrained merger tree realizations as follows. For each galaxy from \cite{Labbe220712446}, we use our estimated halo mass and its reported redshift to set $M_1$ and $z_1$, respectively. We then calculate the most likely distribution of descendant halo masses, given $M_1$ and $z_1$, assuming the halo mass function at $z_0=0$. We derive this descendant mass distribution in Appendix~\ref{sec:descendant_derivation}, finding that the most likely $z=0$ descendant masses for the seven galaxies listed above are $\log (M_{\mathrm{halo}}(z_{0}=0)/\msun)=14.4$, $14.2$, $14.0$, $13.9$, $13.8$, $13.5$, and $13.3$ respectively. We use the full descendant mass distributions derived in Appendix~\ref{sec:descendant_derivation} to draw values of $M_0$ for constrained merger tree realizations; thus, unlike the results presented up to this point, we now use a \emph{distribution} of $M_0$ for each set of $M_1$ and $z_1$ (at fixed $z_0=0$) rather than a single value. Note that, because the JWST galaxy candidates are $\approx 2\sigma$ to $4\sigma$ outliers relative to the underlying halo population, our constrained merger tree realizations are on average $\sim 20$ to $10^4$ times more efficient than a calculation based on down-sampling unconstrained merger trees.

We generate $1000$ realizations of constrained merger trees for each of the seven \cite{Labbe220712446} JWST galaxy candidates in this manner, and Fig.\ \ref{fig:jwst_results} shows the distribution of resulting constrained and main branch growth histories for each system. These growth histories behave similarly to the $M_1=10^{12}~\msun$, $z_1=8$ constraint with $M_0=10^{14}~\msun$ studied above, and typically reach halo masses of $10^{11}~\msun$ (i.e., $10^{-3}$ times the final halo mass) by $z\approx 10$. However, the spread of $M_0$ values predicted by our ePS descendant calculation for each JWST galaxy candidate yields a larger diversity of growth histories than our calculations with a single value of $M_0$; considering the \cite{Labbe220712446} galaxy population as a whole, the spread of constraint parameters derived from the estimated galaxy properties further increases this scatter.

We predict a distribution of half-mass formation times for the main branches of the \cite{Labbe220712446} galaxy candidates with a mean and standard deviation of $z_{1/2}=2.4\pm 1.1$, averaged over the seven systems; this result does not significantly change when we use the constrained branches instead. We also calculate the time, after each system's observation epoch, until the first major merger (defined as any merger with a mass ratio above $1$ to $3$) along the constrained branch that represents each galaxy. For the seven galaxies listed above, we predict that major mergers occur $t_{\mathrm{merger}}=0.8$, $1.4$, $2.0$, $1.8$, $2.5$, $3.3$, and $4.2~\mathrm{Gyr}$ after the corresponding observation epochs, on average. These merger times systematically decrease as the typical final halo mass decreases because the constrained branches then merge onto (or are identical to) lower mass main branches.

Finally, we predict the fates of the high-redshift JWST galaxy candidates by calculating the fraction of these systems predicted to descend to satellites of the main branch halo in their merger trees. In the context of our constrained merger tree realizations, these galaxies are predicted to remain isolated only if the constrained branch and main branches coincide for all $z<z_1$; otherwise, the constrained branch must eventually merge onto a larger halo and becomes a satellite of that system.\footnote{In turn, this requirement implies that the constrained and main branches coincide at all redshifts, because we have defined the constrained branch as the main branch of the tree rooted at $(M_1,z_1)$ for $z>z_1$.} We find that the probability for each \cite{Labbe220712446} galaxy candidate to coincide with its main branch, and thus remain a central galaxy at $z=0$, is $33$, $17$, $17$, $15$, $14$, $7$, and $5$ per cent, respectively. Interestingly, these average isolated fractions decrease as the mean expected $z_0=0$ halo mass decreases. For each galaxy candidate, these isolated fractions are fairly strong functions of $M_0$ and typically drop by a factor of $\sim 2$ when only the upper half of each galaxy candidate’s $M_0$ distribution is considered, because the constrained halo is more likely to merge onto a more massive branch to achieve the final mass in such cases.

Our calculations imply that the JWST galaxy candidates likely become satellites of a larger system at $z\approx 2$, although the most extreme high-redshift systems have a non-negligible chance to remain isolated and inhabit massive ($\gtrsim 10^{14}~\msun$) haloes at $z=0$. Both evolutionary pathways are interesting observationally; mergers at $z\approx 2$ coincide with the peak of the cosmic star formation rate density and may help quench massive clusters at these redshifts. Meanwhile, the fraction of extreme high-redshift systems predicted to remain isolated may be remarkably luminous today, if their star formation continues to be fueled and in the absence of feedback or merger-driven quenching; on the other hand, if they occupy isolated environments, they may quickly deplete their gas and appear as ``fossils'' of the first galaxies today. As discussed below, we plan to study these constrained systems' baryonic components in future work.

\section{The Milky Way--Large Magellanic Cloud Merger}
\label{sec:mw_lmc}

The constrained branches generated using our Brownian bridge excursions do not necessarily coincide with or merge directly onto the main branches of our constrained merger trees. In particular, our constrained solutions only guarantee that the constrained branch reaches a mass of precisely $M_1$ at redshift $z_1$; its position in the overall merger tree is a non-trivial prediction of our model. For example, the constrained halo can be a lower-mass progenitor along the main branch, or it can merge onto a secondary branch that eventually merges with the main branch. Meanwhile, observational constraints on systems' accretion histories are often easiest to interpret as mergers with the main branch progenitor of the main halo. In future work, we therefore plan to consider more general classes of excursion set solutions, and to experiment with directly modifying merger trees during the construction process (e.g., by forcing mergers onto the main branch with desired properties to occur) in order to produce constrained growth histories that match specific sets of observational constraints in detail.

Here, we briefly illustrate that even our constrained Brownian bridge excursions can consistently yield mergers of the constrained halo directly onto the main branch, using the merger between the Milky Way and the Large Magellanic Cloud (LMC) as an example. The LMC is the Milky Way's largest satellite galaxy, with a stellar mass of $3\times 10^9~\msun$ \citep{vanderMarel0205161} and an estimated halo mass of $\approx 1$ to $3\times 10^{11}~\msun$ \citep{Erkal181208192,Erkal190709484,Shipp210713004}. Kinematic measurements imply that the LMC fell into the Milky Way recently, within the last $\approx 2~\mathrm{Gyr}$ \citep{Kallivayalil0508457,Kallivayalil13010832}. This makes the Milky Way--LMC merger fairly rare, in the sense that only~$\approx 10$ per cent of Milky Way-mass haloes host such a massive subhalo today; an even smaller fraction of these massive subhaloes accrete at very late times and reside near the host centre today like the LMC (e.g., \citealt{Liu10112255,Nadler210107810}).

\begin{figure}
    \centering
    \includegraphics[width=0.475\textwidth]{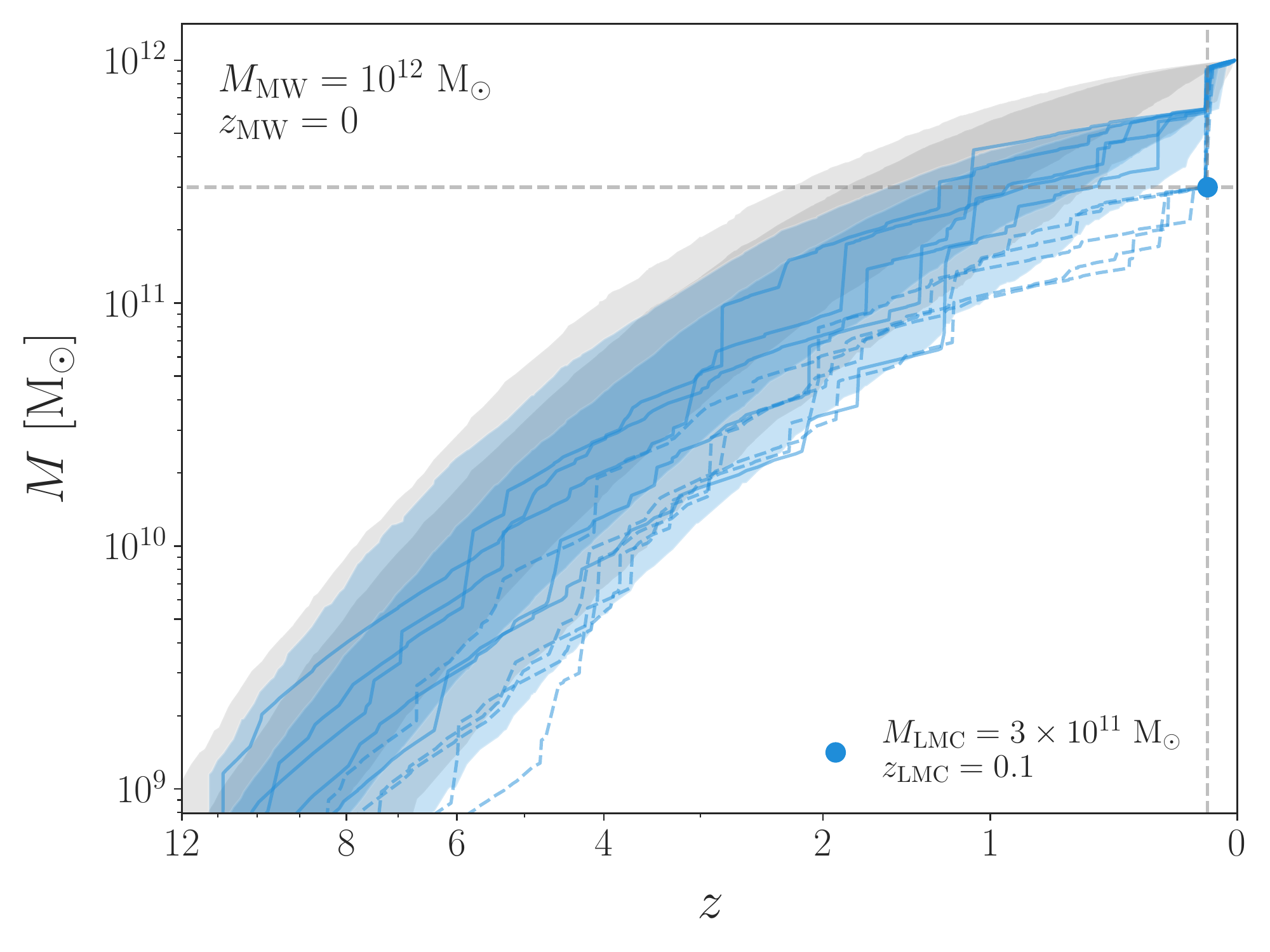}
    \caption{Growth histories for a final halo of mass $M_0=10^{12}~\msun$ at $z_0=0$, similar to the mass of the Milky Way halo. Dark (light) blue bands show $68$ per cent ($95$ per cent) quantiles of the distribution of constrained main branch growth histories for a constraint of $(M_1,z_1)=(3\times 10^{11}~\msun,0.1)$, chosen such that the constrained branch resembles the Large Magellanic Cloud and merges onto the main branch. Five individual constrained growth histories are shown for the main branch (solid lines) and constrained branch (i.e., the Large Magellanic Cloud analogue; dashed lines). Dark (light) grey bands show $68$ per cent ($95$ per cent) quantiles of the distribution of unconstrained main branch growth histories for the same final halo mass and redshift.}
    \label{fig:MW_MAH}
\end{figure}

The Milky Way--LMC merger provides a particularly clean example of a system for which the ``constrained'' branch (i.e., the LMC) should merge directly onto the ``main'' branch (i.e., the Milky Way progenitor at $z\approx 0.1$, corresponding to a lookback time $\approx 1~\mathrm{Gyr}$). In particular, it is observationally disfavoured (and exceedingly unlikely in ePS theory) for a constrained branch that resembles the LMC by reaching a halo mass of $M_1 \approx 3\times 10^{11}~\msun$ at $z_1\approx 0.1$ to merge with a non-main branch progenitor of the Milky Way. This would require another halo even more massive than the LMC to merge with the main branch of the Milky Way host halo at late times, which has a total mass of $M_0\approx 10^{12}~\msun$ at $z_0=0$ (e.g., see \citealt{Bland-Hawthorn160207702} for a review). Thus, we assume that the constrained branch should represent the LMC.

In Fig.\ \ref{fig:MW_MAH}, we show that imposing a constraint of $M_1=3\times 10^{11}~\msun$ at redshift $z_1=0.1$ for a final halo of mass $M_0=10^{12}~\msun$ at $z_0=0$ consistently yields direct mergers of the constrained and main branches that resemble the infall of the LMC into the Milky Way. Interestingly, the main branches of Milky Way progenitors that experience late-time mergers with LMC analogues are biased towards later-forming growth histories relative to the unconstrained distribution at the $\approx 1\sigma$ to $2\sigma$ level. The ability to generate such growth histories without down-sampling represents a first step towards a controlled semi-analytic model of the Milky Way's entire accretion history, including all of its most significant merger events. We intend to pursue this application of our constrained excursion set solutions in a future study.

\section{Potential Caveats}
\label{sec:calibration}

We have shown that our constrained merger trees satisfy the relevant Brownian bridge constraints, and we have described their qualitative and quantitative behaviour. However, we have not attempted to calibrate our predictions for merger rates and halo growth histories to analogous quantities measured in cosmological simulations. Importantly, we do not use the first-crossing distributions derived below to compute (conditional) halo mass functions. Instead, we only use the modified first-crossing \emph{rates} to construct merger trees, as described in~Section~\ref{sec:implementation}. A comparison between the (conditional) population statistics of haloes generated by our constrained solutions and cosmological simulations is beyond the scope of this study and represents an interesting avenue for future work.

There are multiple fronts along which calibration of merger rates and halo growth histories should proceed. First, because we modified standard ePS merger tree construction algorithms as described in Section~\ref{sec:implementation}, and particularly because we sampled progenitor masses from the full interval $[M_{\mathrm{res}},M-M_{\mathrm{res}}]$, even our predictions for \emph{unconstrained} growth histories may require some recalibration. Concretely, such recalibration could follow the procedure in \cite{Benson161001057} by optimizing the parameters of our generalized modifier function (equations (\ref{eq:pch_plus})--(\ref{eq:modifier_symmetrized})) to match merger rates and mass accretion histories measured in cosmological $N$-body simulations; we intend to address this in future work.

Second, it will be necessary to calibrate our predictions for \emph{constrained} growth histories to simulation results. Rather than calibrating to mass accretion histories and merger rates for the full distribution of haloes (e.g., in some interval of final halo mass $M_0$ and redshift $z_0$), this will require calibration to the subset of simulated systems that satisfy a particular constraint (e.g., those with a progenitor that exceeds a mass $M_1$ by redshift $z_1$), performed for many sets of constraint parameters. In practice, this down-sampling procedure may yield very few suitable growth histories in simulations, depending on the simulation size and the extremity of the constraint; for example, haloes with masses comparable to our high-redshift JWST estimates are rare even in large-volume cosmological simulations (e.g., see \citealt{Lovell220810479}). Techniques for ``genetically modifying'' initial conditions to alter the accretion histories of particular haloes found in simulations (e.g., \citealt{Roth150407250,Porciani160900730,Rey170604615,Stopyra200601841}) may aid such comparisons. We anticipate that comparing the correlated nature of halo growth predicted by simulations and by our method in the presence of extreme constraints will be fruitful and may yield new insights into the impact of large-scale environment on smaller scale merger events. We intend to explore these questions in future work.

Finally, we note that, because our implementation of constrained Brownian bridge excursions assumes a constant barrier as a function of $\Delta S$, generalizing our approach for non-constant barriers (e.g., for the warm dark matter barriers considered in \citealt{Benson12093018}) will require dedicated study. 

\section{Discussion and Conclusions}
\label{sec:conclusions}

We have presented a new class of constrained excursion set solutions to model halo growth histories. Specifically, we used Brownian bridges to model constrained excursions of the linear overdensity field in ePS theory. These constrained excursions are guaranteed to pass through a certain overdensity and mass variance, thereby fixing a point in the final halo's growth history, and to accurately represent the distribution of unconstrained excursions that would have satisfied the desired constraint. We implemented these excursions in a modified merger tree construction algorithm to \emph{guarantee} that a progenitor halo with a specific mass exists at a desired redshift, regardless of how rare the resulting growth histories are relative to the underlying, unconstrained distribution.

Our main results focus on merger trees that satisfy extreme high-redshift constraints, and thus represent the tail of the $\Lambda$CDM halo population with the most rapid early growth. Specifically, we have quantified how halo growth histories (Fig.~\ref{fig:mah_constraints}), formation time distributions (Fig.~\ref{fig:formation_time}), and merger rates (Fig.~\ref{fig:deltaM_z}) depend on the constraint parameters. As a proof of principle, we applied our method to predict the most probable halo growth histories of seven high-redshift galaxy candidates identified by \cite{Labbe220712446} in early JWST data (Fig.\ \ref{fig:jwst_results}). These realizations are at least thousands of times more efficient than calculations based on the subset of standard unconstrained merger trees that satisfy the relevant constraints. We predict that the descendants of high-redshift JWST galaxy candidates most likely merge with larger systems $\approx 2~\mathrm{Gyr}$ after the observation epoch and descend to haloes with masses of $\approx 10^{14}~\msun$ today. We will explore whether these systems survive as distinct satellites or merge with the central galaxy in future modelling work that combines semi-analytic galaxy formation models implemented in \textsc{Galacticus} with our constrained solutions.

We have not yet explored how constrained growth histories translate to the \emph{baryonic} components of haloes, e.g., stellar mass, gas mass, and central black hole mass, and the other properties of the galaxies they host. There are significant complexities and uncertainties associated with semi-analytically modelling galaxy formation using ePS merger trees (e.g., see \citealt{Benson10065394} for a review), and we plan to investigate how such astrophysical modelling interplays with our constrained solutions in future studies. Here, we describe some questions that will motivate this work.

First, are the galaxies that occupy rapidly growing haloes that satisfy high-redshift equally significant outliers, relative to the distribution of unconstrained galaxies that exist at the corresponding redshifts, as their dark matter haloes? Naively, the eventual merger of the constrained branch onto the main branch implies that the main halo's accretion rate is increased as a result of the constraint, leading to increased star formation rates and overly luminous central galaxies, compared to unconstrained systems at the same redshift. Several effects may complicate this picture in detail. For example, as we have shown, the galaxy occupying the constrained-branch halo often becomes a satellite of the main-branch progenitor. Depending on the merger time-scale, these constrained-branch satellites can survive as distinct galaxies or merge onto the central; in the case of a major merger between the constrained and main-branch galaxies, star formation can either rapidly cease or be reignited (e.g., see \citealt{Somerville14122712} for a review).

Second, how do (potentially extreme) variations in haloes' growth histories affect the star formation histories of the galaxies they host? As an illustrative example, consider the scatter in the stellar mass--halo mass relation for ultra-faint dwarf galaxies, most of which are thought to be quenched by reionization (e.g., see \citealt{Simon190105465} for a review). Controlled simulations indicate that this scatter is set by how quickly stars formed before reionization. In particular, for a fixed final halo mass, haloes that formed more of their stars before reionization are expected to host more luminous dwarf galaxies \citep{Rey190904664}; this effect may contribute to the large stellar mass--halo mass relation scatter predicted by hydrodynamic simulations at the faint end (e.g., \citealt{Munshi210105822}). In turn, these upscattered systems are expected to accrete more rapidly at early times and can thus be sampled in a controlled manner using our Brownian bridge technique. We can therefore leverage our method -- which enables us to generate thousands of constrained merger trees at the same computational expense as a non-cosmological simulation of comparable resolution -- to explore the relationship between the early growth of low-mass haloes and the resulting properties of the ultra-faint galaxies they host.

Finally, we Note that, supermassive black holes (SMBHs) have been detected at redshifts $z\gtrsim 6$ in optical and near-infrared data (e.g., \citealt{Onoue190407278,Yang200613452}), and these systems are also likely to occupy rare, rapidly growing haloes (e.g., see \citealt{Inayoshi191105791} for a review). Our method can thus be used to study potential connections between early galaxy and SMBH populations in the presence of appropriate seeding mechanisms (e.g., see \citealt{Volonteri221204710} for a recent study).

Our constrained excursion set solutions may in principle be generalized to model the growth histories of astrophysical systems subject to multiple constraints, rather than enforcing a single constraint on the constrained progenitor's mass and redshift. For example, we have already demonstrated that certain regions of our Brownian bridge parameter space consistently yield constrained branches that directly merge onto the main branch, resembling the merger between the Milky Way and Large Magellanic Cloud~(Fig.~\ref{fig:MW_MAH}); we expect that similar techniques can likely reproduce any series of major mergers in a halo's growth history. Enforcing multiple Brownian bridge constraints along a single excursion and implementing other classes of constrained random walks will allow desired growth histories to be reconstructed in increasing detail. In this context, it is interesting to consider the extent to which the detailed growth history of a system like the Milky Way encodes the precise excursion of the linear overdensity field from which it originated.

\section*{Acknowledgements}

We thank Fangzhou Jiang for helpful discussions. Computing resources used in this work were made available by a generous grant from the Ahmanson Foundation. VG amd TD acknowledge the support from the National Science Foundation under grant no. PHY-2013951 and from NASA through the Astrophysics Theory Program, Award Number 21-ATP21-0135.

\section*{Data Availability}

Our constrained merger tree code is publicly available at \url{https://github.com/galacticusorg/galacticus}.


\bibliographystyle{mnras}
\bibliography{references}



\appendix

\onecolumn

\section{Integral Equations for the First-Crossing Distribution}
\label{sec:integral_equations}

\subsection{Unconstrained excursions}
\label{sec:unconstrained_derivation}

For unconstrained excursions, the integral equation that determines $f(S)$ follows from plugging equations (\ref{eq:unconstrained_p}) and (\ref{eq:unconstrained_relative_p}) into equation (\ref{eq:f_integral_final}):
\begin{align}
1 &= \int_0^S f(\hat{S})\mathrm{d}\hat{S} + \int_{-\infty}^{\tilde{B}(S)} \left[\frac{1}{\sqrt{2\pi S}}\exp\left(-\frac{\delta^2}{2S}\right) - \int_{0}^S f(\hat{S})\frac{1}{\sqrt{2\pi (S-\hat{S})}}\exp\left(-\frac{(\delta-\tilde{B}(\hat{S}))^2}{2(S-\hat{S})}\right)\mathrm{d}\hat{S}\right]\mathrm{d}\delta & \nonumber \\ &= \int_0^S f(\hat{S})\mathrm{d}\hat{S} + \mathrm{erf}\left[\frac{\tilde{B}(S)}{\sqrt{2S}}\right] - \int_0^S f(\hat{S}) \mathrm{erf}\left[\frac{\tilde{B}(S)-\tilde{B}(\hat{S})}{\sqrt{2(S-\hat{S})}}\right]\mathrm{d}\hat{S}. &\label{eq:f_integral_unconstrained_full}
\end{align}
For a constant barrier $\tilde{B}(S)\equiv B$ (relevant, for example, in ePS predictions for cold dark matter cosmologies), the final term vanishes because $\tilde{B}(S)-\tilde{B}(\hat{S})=0$. Equation (\ref{eq:f_integral_unconstrained_full}) can then be solved by differentiating with respect to $S$, which yields the standard first-crossing distribution
\begin{equation}
    f_{\mathrm{constant\ barrier}}(S)=\frac{B}{\sqrt{2\pi S^3}}\exp\left(-\frac{B^2}{2S}\right).
\end{equation}

\subsection{Constrained Brownian bridge excursions}
\label{sec:constrained_derivation}

For our constrained Brownian bridge excursions, the the integral equation that determines $f(S)$ follows from plugging equations (\ref{eq:constrained_p}) and \ref{eq:constrained_relative_p} into equation (\ref{eq:f_integral_final}). To write this explicitly, we first expand equation (\ref{eq:constrained_p}) and equation (\ref{eq:constrained_relative_p}) according to
\begin{align}
P'_0(\delta,S) &= \mathcal{N}(\Delta \delta',\Delta S')& \nonumber \\  &= \mathcal{N}\left(\delta - \tilde{B}(\hat{S}) - \frac{S-S_0}{S_1-S_0}(\delta_1-\delta_0),\frac{(S_1-S)(S-S_0)}{S_1-S_0}\right) & \nonumber \\ &= \frac{1}{\sqrt{2\pi}}\sqrt{\frac{(S_1-S_0)}{ (S_1-S)(S-S_0)}}\exp\left(-\frac{\left[\delta - \tilde{B}(\hat{S}) - \frac{S-S_0}{S_1-S_0}(\delta_1-\delta_0)\right]^2}{2\frac{(S_1-S)(S-S_0)}{(S_1-S_0)}}\right),&
\end{align}
\begin{align}
P'_0[\delta,S|\tilde{B}(\hat{S}),\hat{S}] &= \mathcal{N}(\Delta \delta',\Delta S') & \nonumber \\ &= \mathcal{N}\left(\delta - \tilde{B}(\hat{S}) - \frac{S-\hat{S}}{S_1-S_0}(\delta_1-\delta_0),\frac{(S_1-S)(S-\hat{S})}{S_1-S_0}\right)& \nonumber \\ &= \frac{1}{\sqrt{2\pi}}\sqrt{\frac{(S_1-S_0)}{ (S_1-S)(S-\hat{S})}}\exp\left(-\frac{\left[\delta - \tilde{B}(\hat{S}) - \frac{S-\hat{S}}{S_1-S_0}(\delta_1-\delta_0)\right]^2}{2\frac{(S_1-S)(S-\hat{S})}{(S_1-S_0)}}\right).&
\end{align}
Finally, we plug these expressions into equation (\ref{eq:f_integral_final}) to obtain
\begin{align}
1 &= \int_0^S f(\hat{S})\mathrm{d}\hat{S} + \int_{-\infty}^{\tilde{B}(S)} \Bigg[\frac{1}{\sqrt{2\pi}}\sqrt{\frac{(S_1-S_0)}{ (S_1-S)(S-S_0)}}\exp\left(-\frac{\left[\delta - \tilde{B}(\hat{S}) - \frac{S-S_0}{S_1-S_0}(\delta_1-\delta_0)\right]^2}{2\frac{(S_1-S)(S-S_0)}{(S_1-S_0)}}\right) & \nonumber \\ &\hspace{35mm} - \int_{0}^S f(\hat{S})\frac{1}{\sqrt{2\pi}}\sqrt{\frac{(S_1-S_0)}{ (S_1-S)(S-\hat{S})}}\exp\left(-\frac{\left[\delta - \tilde{B}(\hat{S}) - \frac{S-\hat{S}}{S_1-S_0}(\delta_1-\delta_0)\right]^2}{2\frac{(S_1-S)(S-\hat{S})}{(S_1-S_0)}}\right)\mathrm{d}\hat{S}\Bigg]\mathrm{d}\delta & \nonumber \\ &= \int_0^S f(\hat{S})\mathrm{d}\hat{S} + \mathrm{erf}\left[\frac{\tilde{B}(S)-\tilde{B}(\hat{S}) - \frac{S-S_0}{S_1-S_0}(\delta_1-\delta_0)}{\sqrt{2\frac{(S_1-S)(S-\hat{S})}{(S_1-S_0)}}}\right] - \int_0^S f(\hat{S}) \mathrm{erf}\left[\frac{\tilde{B}(S)-\tilde{B}(\hat{S})-\frac{S-\hat{S}}{S_1-S_0}(\delta_1-\delta_0)}{\sqrt{2\frac{(S_1-S)(S-\hat{S})}{(S_1-S_0)}}}\right]\mathrm{d}\hat{S}. &\label{eq:f_integral_constrained_full}
\end{align}
The arguments of the error functions in equation (\ref{eq:f_integral_constrained_full}) simplify for a constant barrier, but even in this case there is no analytic solution for $f(S)$ in our constrained model because the argument of the error function inside the last integral depends on $S$.

\subsection{Numerical Solution Technique}
\label{sec:numerical_solution}

To solve equations (\ref{eq:f_integral_unconstrained_full}) and (\ref{eq:f_integral_constrained_full}) for the unconstrained and constrained first-crossing distributions, respectively, we follow \cite{Benson12093018} by discretizing variance into $N$ logarithmically spaced intervals of width $\Delta S$. We then use the midpoint integration method described in \cite{Du160820575} to discretize the integrals; \cite{Du160820575} show that this procedure outperforms previous solutions based on trapezoid integration. The solution is iterative: at each step, $f(S_i)$ only depends on all $f(S_j)$ (where $j<i$), and on the barrier. We implement this solution generally, such that the first-crossing distribution can be obtained for any (non-constant) barrier in both unconstrained and constrained cases.

\section{Convergence Tests of Constrained Solutions}
\label{sec:convergence_tests}

\subsection{Convergence to unconstrained solutions}

\begin{figure}
    \centering
    \includegraphics[width=0.455\textwidth]{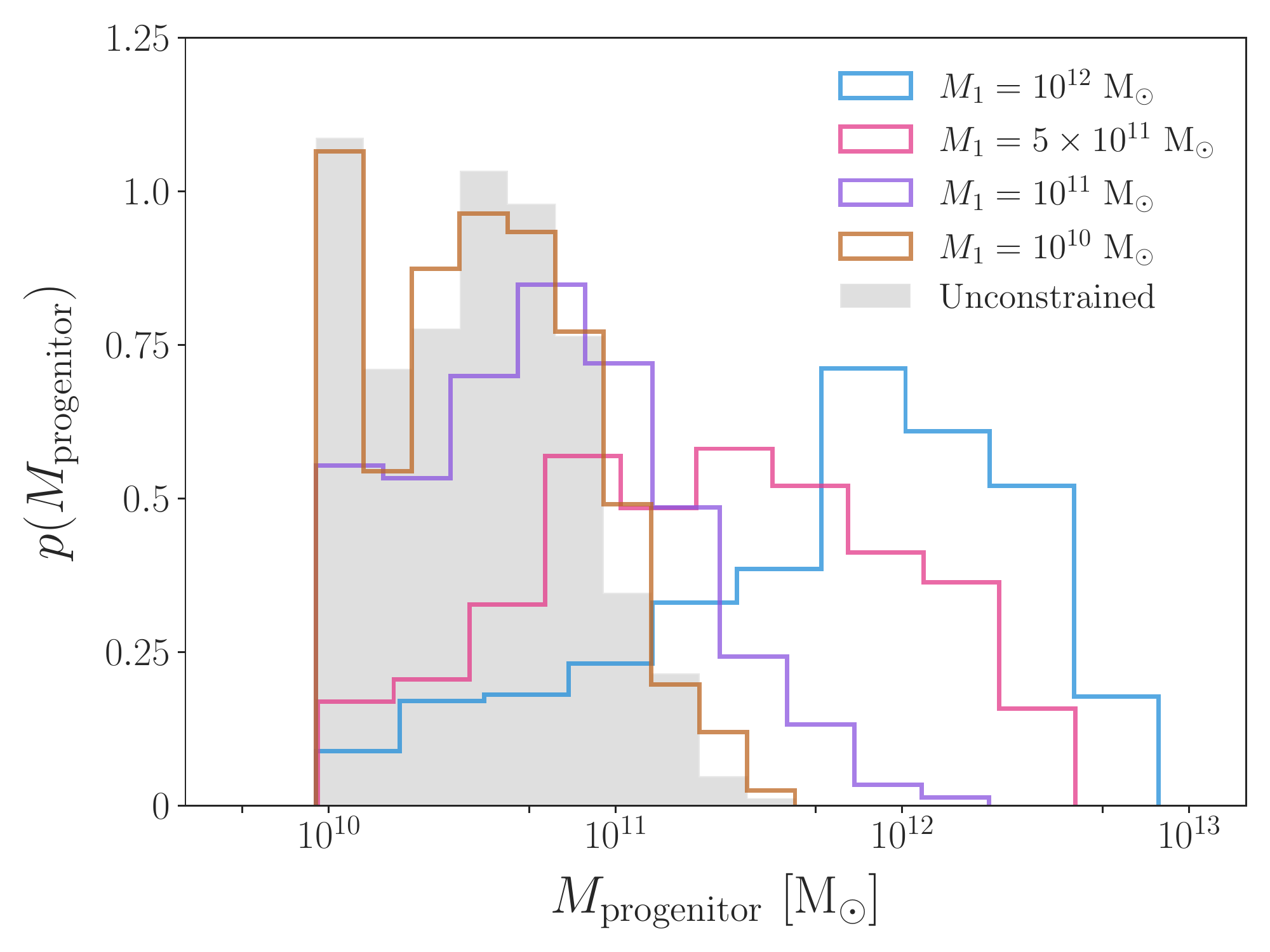}
    \hspace{0.25cm}
    \includegraphics[width=0.5\textwidth]{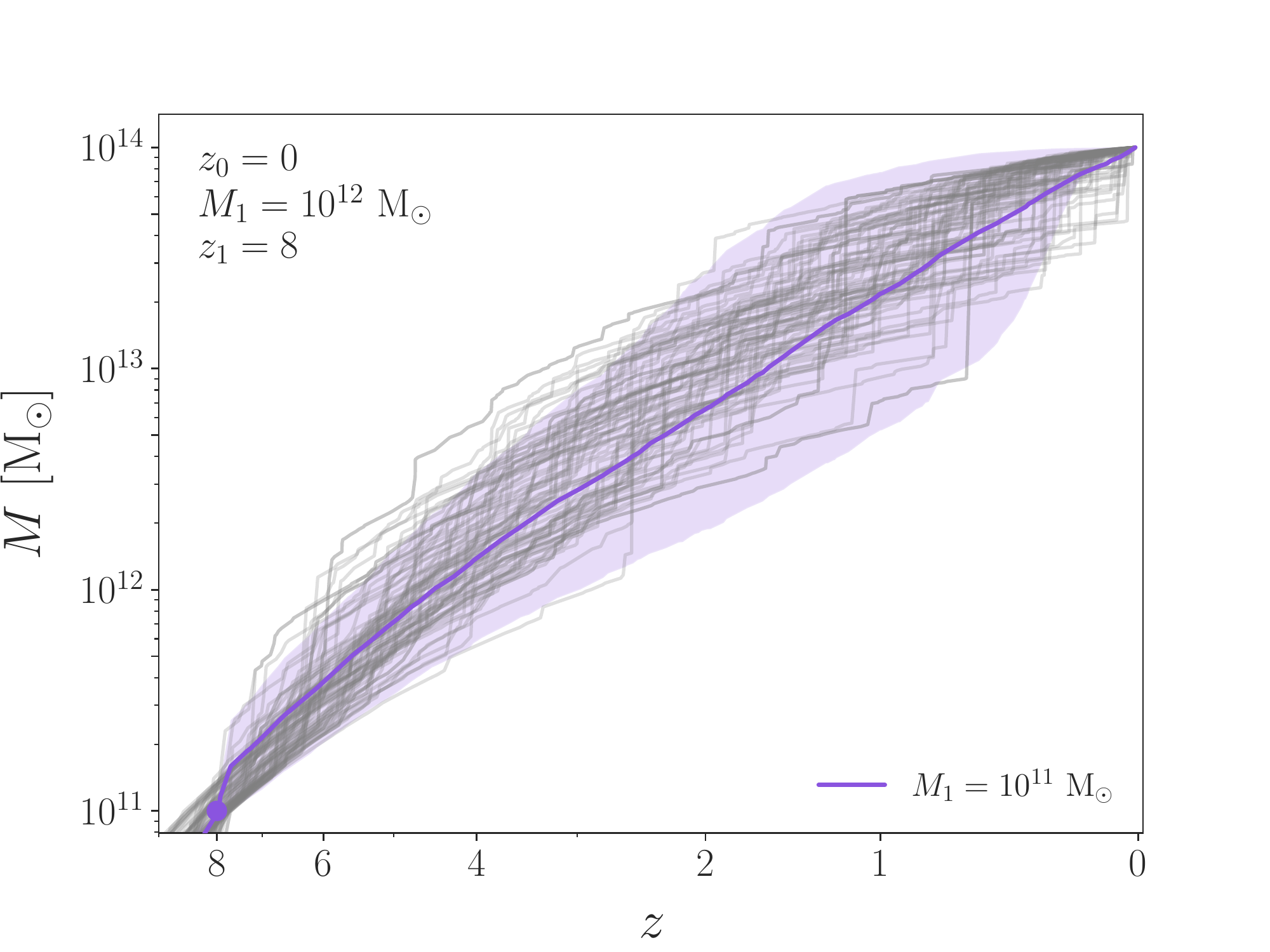}
    \caption{\emph{Left}: The distribution of main branch progenitor masses at $z=8$ for a final halo of mass $M_0=10^{14}~\msun$ at $z_0=0$. Results are shown for $1000$ realizations of constrained merger trees with $M_1=10^{12}~\msun$ (blue), $M_1=5\times 10^{11}~\msun$ (pink), $M_1=10^{11}~\msun$ (purple), and $M_1=10^{10}~\msun$ (brown), all with $z_1=8$, and for unconstrained merger trees with the same final halo properties (grey). \emph{Right}: Comparison of growth histories for constrained branches from constrained merger tree realizations with $z_1=8$ and $M_1=10^{11}~\msun$ (purple) versus branches of unconstrained trees that satisfy this constraint (grey).}
    \label{fig:M1_convergence}
\end{figure}

As argued in Section~\ref{sec:contrained_solution} and indicated in the top-left-hand panel of Fig.~\ref{fig:mah_constraints}, our constrained halo growth histories should converge to unconstrained distributions generated (using the same final halo mass and redshift) for appropriate choices of the constraint parameters. In particular, in the limit $M_1\rightarrow 0$ ($S_1\rightarrow \infty$), our Brownian bridge excursions precisely converge to unconstrained excursions.\footnote{We Note that, convergence properties are more subtle as a function of $\delta_1$. As $\delta_1\rightarrow \delta_0$, the Brownian bridge drift term (equation (\ref{eq:drift_brownian_bridge})) vanishes, but the covariance (equation (\ref{eq:cov_brownian_bridge})) differs from the unconstrained case. This follows because unconstrained trajectories may not pass through $(S_1,\delta_0)$ unless $S_1$ is sufficiently large. Thus, although the growth histories in the top-right-hand panel of Fig.~\ref{fig:mah_constraints} approach the unconstrained distribution as $z_1$ decreases, convergence as a function of $z_1$ (or $\delta_1$) requires a sufficiently small $M_1$ (or large $S_1$).} We quantify how growth histories converge to the unconstrained case by comparing main branch progenitor halo mass distributions as a function of $M_1$, at fixed $M_0$, $z_0$, and $z_1$, at several redshifts $z$. The left-hand panel of Fig.\ \ref{fig:M1_convergence} shows this result at $z=8$ for merger trees constructed using a final halo mass of $M_0=10^{14}~\msun$ at $z_0=0$, with $z_1=8$, and demonstrates the desired behaviour. In particular, two-sample Kolmogorov–Smirnov (KS) tests with the unconstrained distribution yield KS statistics of $0.77$, $0.6$, $0.25$, and $0.05$, for $M_1=10^{12}$, $5\times 10^{11}$, $10^{11}$, and~$10^{10}~\msun$, respectively. The $M_1=10^{10}~\msun$ case is consistent with the unconstrained distribution; in all other cases, the KS test rejects the null hypothesis of distinguishable distributions with $p\ll 0.05$. We repeat these tests for various redshifts and constraint parameters, finding that constrained merger trees with~$M_1\lesssim 10^{-4}M_0$ yield main branch progenitor mass distributions consistent with corresponding unconstrained cases.

We also test whether the subset of unconstrained merger tree realizations that satisfy a given constraint have growth histories that are consistent with those generated directly by our constrained solutions. The right-hand panel of Fig.\ \ref{fig:M1_convergence} demonstrates that the growth histories of specific branches of unconstrained trees that satisfy a $z_1=8$, $M_1=10^{11}~\msun$ constraint agree well with the distribution of constrained branches from our constrained merger trees. We find similar agreement for all constraint parameters studied in this work; thus, our constrained solutions accurately sample the underlying unconstrained distribution as desired.

\subsection{Convergence as a function of numerical resolution}

To test for convergence as a function of the merger tree mass resolution, $M_{\mathrm{res}}$, we repeat the tests from the previous subsection using three sets of merger tree realizations with the same constraint parameters ($M_1=10^{13}~\msun$ and $z_1=8$) and final halo parameters ($M_0=10^{14}~\msun$ and $z_0=0$), with $M_{\mathrm{res}}=10^{11}~\msun$, $10^{10}~\msun$, and $10^9~\msun$, respectively, at several redshifts. We find that the $M_{\mathrm{res}}=10^{11}~\msun$ progenitor mass distribution is inconsistent with our fiducial $M_{\mathrm{res}}=10^{10}~\msun$ results, with $p<0.05$ from KS two-sample tests at all redshifts $1<z\leq 8$ (for $z\lesssim 1$, the distributions agree by construction because the merger trees all reach the same final halo mass). Meanwhile, the $M_{\mathrm{res}}=10^9~\msun$ progenitor mass distributions are consistent with the $M_{\mathrm{res}}=10^{10}~\msun$ case at all redshifts we test, with KS statistics $\leq 0.1$, implying that our fiducial results are well converged for these constraint parameters. We repeat these tests for all constraints used in this paper, finding good convergence for $M_{\mathrm{res}}\leq10^{-4}M_0$ in all cases. We Note that, our constraints also introduce another convergence criterion -- namely, that $M_{\mathrm{res}}\leq M_1$ so that the constrained-branch merger is always resolved. In practice, we find that $M_{\mathrm{res}}\leq10^{-1}M_1$ is sufficient to achieve converged constrained-branch progenitor mass distributions. We have also tested the effects of merger tree construction resolution parameters in \textsc{Galacticus} that control the resolution of the first-crossing rate distribution as a function of variance and time. None of our results are sensitive to variations of our fiducial choices for these parameters, which are documented in our publicly available code.

\pagebreak

\section{Descendant Halo Mass Distributions}
\label{sec:descendant_derivation}

In our constrained solutions, we can view the final halo mass $M_0$ and redshift $z_0$ as free parameters, given a constraint that the merger tree exceeds a mass of $M_1$ at some redshift $z_1$. Thus, for the high-redshift constraints considered in this work, it is natural to consider what distribution of halo masses, $p(M_0,z_0)$, these haloes descend to at redshift $z_0$, assuming they self-consistently evolve to produce the halo mass function at $z=z_0$ calculated from standard unconstrained ePS solutions. In the context of our excursion set solutions, the probability that a trajectory which passes through $(S_1,\delta_1)$ will descend to a halo of mass $S_0$ at some fixed $\delta_0$ is
\begin{equation}
    p(S_0|\delta_0,\delta_1,S_1) \propto p(\delta_1|S_0,S_1,\delta_0)p(S_0,\delta_0) \propto p(\delta_1|S_0,S_1,\delta_0)\frac{\mathrm{d}n(S_0,\delta_0)}{\mathrm{d}S_0} \propto p(\delta_1|S_0,S_1,\delta_0)\frac{\mathrm{d}n(M_0[S_0],z_0(\delta_0))}{\mathrm{d}M_0}\left|\frac{\mathrm{d}M_0[S_0]}{\mathrm{d}S_0} \right|,\label{eq:descendant_prob}
\end{equation}
where $p(\delta_1|S_0,S_1,\delta_0)\mathrm{d}\delta_1$ is the probability that a trajectory starting from $S_0$ (at some fixed $\delta_0$) will pass through a small region $(S_1,\delta_1\pm \mathrm{d}\delta_1)$, $\mathrm{d}n(M_0[S_0],z_0(\delta_0))/\mathrm{d}M_0$ is the halo mass function evaluated at the redshift $z_0$ corresponding to $\delta_0$, and the Jacobian $\left|\mathrm{d}M_0[S_0]/\mathrm{d}S_0 \right|$ follows from equation (\ref{eq:root-variance}). \cite{Lacey1993} provide an expression for $p(\delta_1|S_0,S_1,\delta_0)\mathrm{d}\delta_1$, assuming that \emph{any} branch of the merger tree can satisfy the constraint (specifically, these are the ``constrained'' branches in our implementation). In our notation, their equation (2.16) becomes
\begin{equation}
    p_{\mathrm{constrained\ branch}}(\delta_1|S_0,S_1,\delta_0) \propto \frac{1}{\sqrt{2\pi}}\left[\frac{S_1}{S_0(S_1-S_0)}\right]^{3/2}\frac{\delta_0(\delta_1-\delta_0)}{\delta_1}\times \exp\left[-\frac{(\delta_0 S_1-\delta_1 S_0)^2}{2S_1S_0(S_1-S_0)}\right],\label{eq:constrained_prob}
\end{equation}
for $S_1>S_0$ and $\delta_1>\delta_0$, and zero otherwise. We also calculate the probability that a \emph{single} branch of the merger tree (which we assume to be the main branch) satisfies the constraint and descends to $(S_0,\delta_0)$. Since any single (unconstrained) trajectory is an uncorrelated random walk with Gaussian transition probabilities, we have
\begin{equation}
   p_{\mathrm{main\ branch}}(\delta_1|S_0,S_1,\delta_0) \propto \frac{1}{\sqrt{2\pi}} \frac{1}{\sqrt{S_1-S_0}} \times \exp\left[-\frac{(\delta_1-\delta_0)^2}{2(S_1-S_0)}\right],\label{eq:main_prob}
\end{equation}
which again holds for $S_1>S_0$ and $\delta_1>\delta_0$ and is zero otherwise. Note that, this expression can also be interpreted in terms of the conditional progenitor mass function for the final halo.

\begin{figure}
    \centering
    \includegraphics[width=0.475\textwidth]{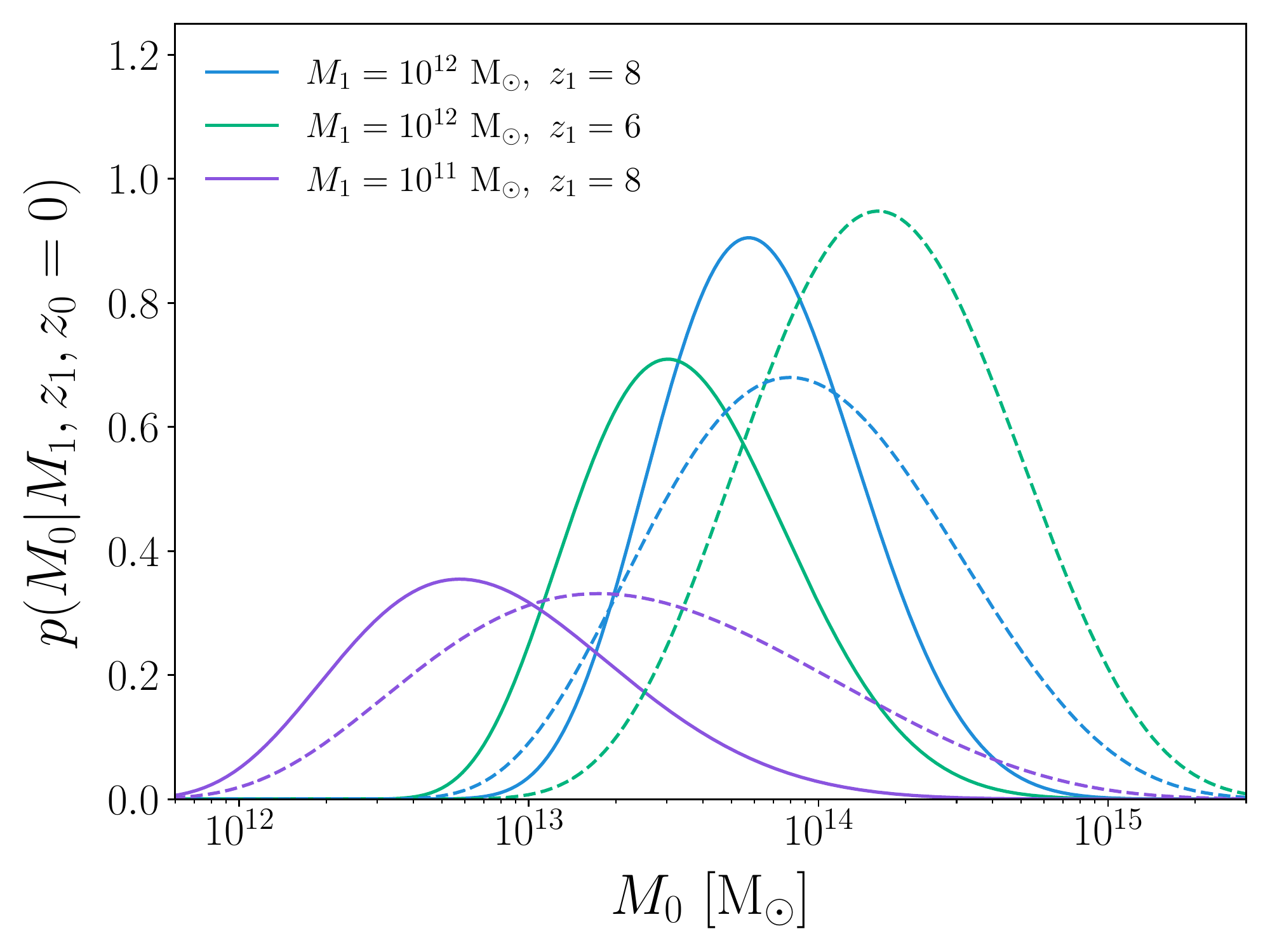}
    \includegraphics[width=0.475\textwidth]{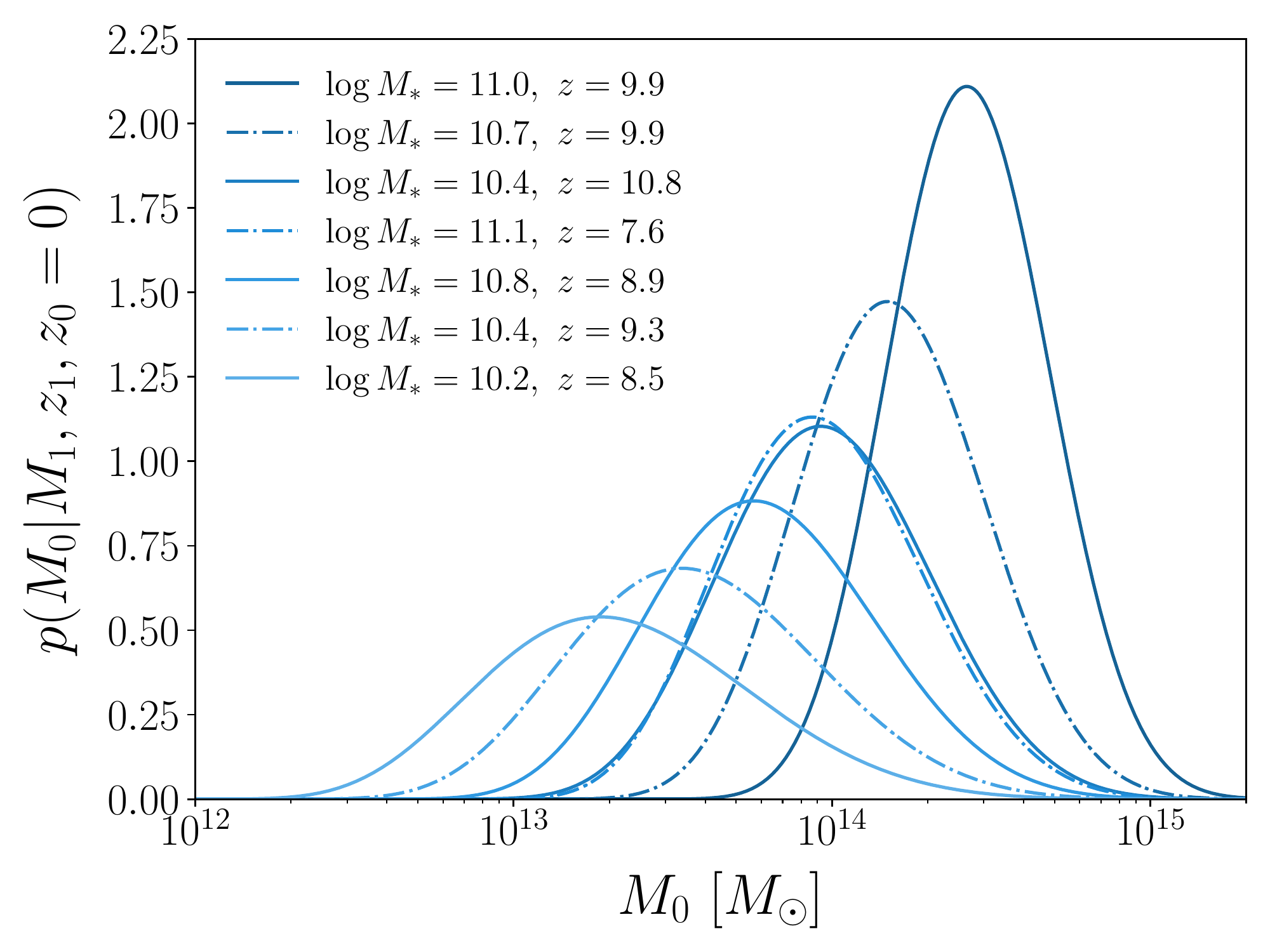}
    \caption{\emph{Left}: The distribution of $z_0=0$ descendant halo masses for excursions that exceed a mass of $M_1=10^{12}~\msun$ at $z_1=8$ (blue), $M_1=10^{12}~\msun$ at $z_1=6$ (green), and $M_1=10^{11}~\msun$ at $z_1=8$ (purple). Solid (dashed) lines correspond to the case where any branch (a single branch) satisfies the constraint. \emph{Right}: Same as the left-hand panel, for the seven high-redshift galaxy candidates from \protect\cite{Labbe220712446}, using the progenitor halo masses estimated in Section~\ref{sec:highz_halo_mass}.
    \label{fig:M1_distribution}}
\end{figure}

We calculate the constrained and main-branch forms of the descendant distribution by assuming $z_0=0$ and plugging equations (\ref{eq:constrained_prob}) and (\ref{eq:main_prob}) into equation (\ref{eq:descendant_prob}), respectively, using a $z_0=0$ halo mass function computed from our unconstrained solutions. We then change variables from $(S,\delta)$ to $(M,z)$ by tabulating numerical solutions for the mass--variance relation and halo mass function. Fig.\ \ref{fig:M1_distribution} shows the resulting descendant halo mass distributions for several constraints considered throughout this work. These distributions are suppressed at both high- and low-descendant masses. At high descendant masses, the $z_0=0$ halo mass function cuts off exponentially, while at low-descendant masses, the probability of a growth history exceeding the desired mass by redshift $z_1$ is also exponentially suppressed; for $M_0<M_1$, the probability vanishes. Note that, the constrained-branch distribution is shifted towards larger descendant halo masses relative to the main-branch distribution. This follows because a specific branch that exceeds the high-redshift constraint continues growing for $z<z_1$, while \emph{any} branch that satisfies the constraint can later merge onto the main branch. The results in Fig.\ \ref{fig:M1_distribution} imply that a halo that exceeds mass $M_1=10^{12}~\msun$ ($10^{11}~\msun$) at redshift $z_1=8$ most likely descends to a halo of mass $M_0=6\times 10^{13}~\msun$ ($6\times 10^{12}~\msun$) at $z_0=0$. Decreasing the constraint redshift at fixed progenitor mass also shifts the $z_0=0$ descendant distribution towards lower masses, as expected. For example, in the ($M_1=10^{12}~\msun$ $z_1=6$) case, the most likely descendant halo mass is $M_0=3\times 10^{13}~\msun$.

\bsp	
\label{lastpage}
\end{document}